%% file: paper_ieee.tex
\documentclass[conference]{IEEEtran}

\input{preamble.tex}

\title{\textsc{RepTran}: Search-Based Repair of\\Transformer Models}

\input{author_info_ieee.tex}

\IEEEtitleabstractindextext{%
\begin{abstract}
\input{section/abstract.tex}
\end{abstract}

\begin{IEEEkeywords}
AI-enabled software, Transformer, Repair, Reliability
\end{IEEEkeywords}
}

\begin{document}
\maketitle
% compsoc のときは次の行で abstract/keywords を表出
\IEEEdisplaynontitleabstractindextext

% ---------------------------- Main Body ----------------------------
\input{section/intro.tex}

\input{section/background.tex}
\input{section/method.tex}
\input{section/exp_setup.tex}
\input{section/evaluation.tex}
\input{section/discussion.tex}
\input{section/validity.tex}
\input{section/conclusion.tex}
\input{section/data_availability.tex}

% ---------------------------- Acknowledgments ----------------------------
\section*{Acknowledgments}
We gratefully acknowledge the financial support of:
(1) Japan Society for the Promotion of Science (JSPS) for the KAKENHI grants (JP25K22845 and JP26H02500);
(2) Japan Science and Technology Agency (JST) as part of Adopting Sustainable Partnerships for Innovative Research Ecosystem (ASPIRE), Grant Numbers JPMJAP2415 and JPMJAP2301;
(3) JST-Mirai Program Grant No.\ JPMJMI20B8;
(4) the Inamori Research Institute for Science (InaRIS Fellowship to Yasutaka Kamei).

% ---------------------------- References ----------------------------
% IEEE は番号スタイル。bst は IEEEtran を使用
\bibliographystyle{IEEEtran}
\bibliography{IEEEabrv, reference}

\end{document}

%% file: preamble.tex
\AtBeginDocument{%
  \providecommand\BibTeX{{%
    \normalfont B\kern-0.5em{\scshape i\kern-0.25em b}\kern-0.8em\TeX}}}

\usepackage{cite}% 通常（[1]–[3] などの圧縮あり）
\usepackage{amsmath,amssymb,amsfonts}

\usepackage{graphicx}      % 画像
\usepackage{booktabs}      % 表の罫線
\usepackage{multirow}
\usepackage{makecell}
\usepackage[table]{xcolor} % 表セルの色

\usepackage[caption=false,font=footnotesize]{subfig}

\usepackage{algorithm}
\usepackage{algorithmic}
\usepackage{enumitem}

\usepackage{float}
\usepackage{wrapfig}
\usepackage{comment}
\usepackage{threeparttable}
\usepackage{hhline}
\usepackage{xspace}
\usepackage{soul}      % \ul など
\usepackage{xcolor}
\usepackage{framed}
\usepackage[skins,breakable]{tcolorbox}
\tcbuselibrary{raster,skins}
\usepackage{url}       % URL（hyperref と併用OK）
\usepackage[normalem]{ulem}

\def\figref#1{Fig.~\ref{#1}}
\def\tabref#1{Tab.~\ref{#1}}
\def\secref#1{Sect.~\ref{#1}}
\def\et{et\ al.\xspace}
\def\eg{e.g.\xspace}
\def\ie{i.e.\xspace}
\def\reptran{\textsc{RepTran}\xspace}
\def\arachne{\textsc{Arachne}\xspace}
\def\arachnew{\textsc{ArachneW}\xspace}

\definecolor{mygray}{gray}{0.9}
\definecolor{ForestGreen}{HTML}{288c66}
\definecolor{MyBlue}{HTML}{00008b}
\setulcolor{MyBlue}
\definecolor{gold}{HTML}{FFD700}
\definecolor{silver}{HTML}{C0C0C0}
\definecolor{bronze}{HTML}{B36B00}

\def\citemodelrepair{\cite{sun2022ICSE,sohn2022tosem,calsi2023ICST,li2023GECCO,tao2023PLDI,sun2024tosem,nawas2024isaiv,you2025arxiv}\xspace}

\def\citednntesting{\cite{guo2025tosem, yuheng2025tse}\xspace}

\newtcolorbox{summaryboxenv}{%
  colback=gray!10,
  colframe=black,
  boxrule=0.4pt,
  arc=0pt,
  outer arc=0pt,
  left=1.5pt, right=1.5pt, top=1.5pt, bottom=1.5pt,
  boxsep=0pt,
}
\newcommand{\summaryblock}[2]{%
{\vspace{-6pt}%
\begin{summaryboxenv}
\noindent \textbf{#1:} #2
\end{summaryboxenv}%
\vspace{-9pt}%
}}

\usepackage[hidelinks]{hyperref}

%% file: author_info_ieee.tex
\author{
    \IEEEauthorblockN{
        Yuta Ishimoto\textsuperscript{*},
        Paolo Arcaini\textsuperscript{\dag},
        Fuyuki Ishikawa\textsuperscript{\dag},
        Masanari Kondo\textsuperscript{\ddag},
        Naoyasu Ubayashi\textsuperscript{\S},
        and Yasutaka Kamei\textsuperscript{\ddag}
    }
    \IEEEauthorblockA{
        \textsuperscript{*}The University of Osaka, Japan,
        \textsuperscript{\dag}National Institute of Informatics, Japan,\\
        \textsuperscript{\ddag}Kyushu University, Japan,
        \textsuperscript{\S}Waseda University, Japan
    }
}

%% file: section/abstract.tex
% Context
To ensure the overall quality of AI-enabled software, not only traditional software components but also AI components need to be tested and repaired.
Among AI components, Transformer models are increasingly integrated into software systems, which makes their misbehaviors critical.
Although prior work in the software engineering community has proposed deep neural network (DNN) repair methods, most overlook Transformer-specific structures.
% Objective
We propose \reptran, a search-based repair method for Transformer models.
It targets their feed-forward networks (FFNs), which play a central role in the architecture.
\reptran identifies suspicious weights by combining two types of scores: a variance-based neuron score and an existing bidirectional score.
It then iteratively optimizes these weights using differential evolution.
% Results
Our evaluation includes 18 fault benchmarks constructed from CIFAR-100 and Tiny-ImageNet.
We compare \reptran against three baselines: random weight selection, \arachne (a state-of-the-art DNN repair method), and \arachnew, which enables \arachne to control the number of selected weights.
\reptran achieved an average repair rate of 74.7\%, statistically outperforming random selection and \arachne across all benchmarks.
Effect size analysis revealed that \reptran achieved higher repair rates than \arachnew regardless of the number of selected weights.
% Conclusion
These results suggest that \reptran is effective for enhancing the reliability of AI-enabled software.

%% file: section/intro.tex
\section{Introduction} \label{sec:intro}
The Transformer~\cite{vaswani2017NIPS} has become a critical neural network architecture underlying foundation models such as large language models (LLMs).
% Originally proposed for natural language processing (NLP), the Transformer has been adapted to a wide variety of tasks (\eg, computer vision (CV)~\cite{dosovitskiy2021ICLR, khan2022csur}).
Since Transformer models are increasingly used in a wide range of AI-enabled software (\eg, autonomous driving~\cite{hu2023cvpr}), misbehaviors of these models can significantly compromise the reliability of the entire system.
% \red{
% These risks have already materialized in practice:
% for example, the OECD AI Incidents and Hazards Monitor~\cite{oecd_aim} records a case in which Tesla’s Autopilot was found partly liable for a fatal crash, leading to a jury verdict of up to \$329 million in damages.\footnote{\url{https://www.bbc.com/news/articles/c93dqpkwx4xo}, accessed: 2025-09-02.}
% }
Indeed, the OECD AI Incidents and Hazards Monitor has documented numerous cases of failures in autonomous driving systems~\cite{oecd_aim_av}, many of which resulted in severe injuries, fatalities, and significant financial losses.
% For example, Tesla’s Autopilot was found partly liable for a fatal crash, leading to a jury verdict of up to \$329 million in damages.\footnote{\url{https://www.bbc.com/news/articles/c93dqpkwx4xo}, accessed: 2025-09-02.}

One promising direction for addressing such misbehaviors is \textit{deep neural network (DNN) repair}~\citemodelrepair.
% , which has been actively studied in the software engineering (SE) community~\citemodelrepair.
These methods are designed to complement full retraining:
while full retraining aims to improve overall performance (\eg, accuracy), DNN repair targets specific types of misbehavior.
It identifies and updates only the components of the model (\eg, neurons or weights) that are responsible for the misbehavior.
This selective update seeks to correct the targeted misbehavior while preserving the existing correct behaviors.

Given this background, there is a growing need for repair methods tailored to Transformer models;
however, most existing DNN repair methods have focused on conventional networks (e.g., convolutional neural networks (CNNs)~\cite{sohn2022tosem,li2023GECCO,calsi2023ICST,sun2024tosem}), with limited attention to Transformer models~\cite{nawas2024isaiv}.
Unlike such conventional networks, Transformer models rely on unique architectural components such as multi-head self-attention (MSA) and feed-forward networks (FFNs) consisting of two fully connected layers with a non-linear activation in between~\cite{vaswani2017NIPS}.
Existing methods do not account for these components, indicating the need for repair methods designed for Transformer models.

\begin{comment}
    To develop a repair method for Transformer models, we can draw on findings from the artificial intelligence (AI) community.
    % The community has paid particular attention to the role of FFNs in Transformer models~\cite{geva2021emnlp,dai2022ACL,kobayashi2020EMNLP}.
    For instance, several studies suggest that intermediate neurons in FFNs may store knowledge learned by the model, motivating efforts to identify \textit{knowledge neurons}~\cite{dai2022ACL}.
    In contrast, MSA has been shown to be more redundant, with some heads being prunable without significant accuracy degradation~\cite{michel2019nips,li2023tvcg}.
\end{comment}
    
% Building on these insights,
To this end, we propose \reptran, a novel repair approach for Transformer models.
% It selectively modifies weights within FFNs to correct specific misbehaviors while preserving correct behaviors.
% Restricting repair operations to FFNs helps reduce the number of parameters involved.
To identify which weights should be modified, \reptran computes a \textit{weight suspiciousness score}, which reflects the contribution of each weight to the misbehavior.
This score consists of two components: (1) a \textit{neuron score} and (2) a \textit{bidirectional score}.
The neuron score, newly introduced in this work, is based on the mean and variance of activations of intermediate FFN neurons.
% By comparing activations from correctly and incorrectly classified samples, it highlights neurons whose behaviors differ the most.
% This score is inspired by neural coverage (NLC)~\cite{yuan2023ICSE}, the latest DNN coverage metrics.
The bidirectional score comes from \arachne~\cite{sohn2022tosem}, a state-of-the-art DNN repair method.
% It measures how much each weight affects the model output in both forward and backward directions.
% The weight suspiciousness score is then computed by combining the neuron score and bidirectional score.
% This combination enables FFN-aware suspicious weight selection for Transformer models that captures both structural and behavioral characteristics of these models.
Weights with high suspiciousness scores are selected for repair.
These weights are optimized using differential evolution (DE)~\cite{storn1997de}, a population-based search algorithm.
The fitness function is designed to maximize the number of repaired misbehaviors of the target type while minimizing disruptions to the existing correct behaviors.

% exp. setting
We evaluated \reptran on 18 fault benchmarks constructed from two image classification datasets: CIFAR-100~\cite{krizhevsky2009learning} and Tiny-ImageNet~\cite{le2015tImg}.
% These benchmarks include not only the source-to-target (\textit{SRC-TGT}) misclassifications (\ie, misclassifying samples from class A as class B), but also missed detections for a specific target class (\textit{TGT-FN}) and false detections as a target class (\textit{TGT-FP}), reflecting real-world scenarios where deployed AI-enabled systems are hindered by high false positives or false negatives.
We used the Vision Transformer (ViT)~\cite{dosovitskiy2021ICLR} as the target model in our experiments.
% exp. results
\reptran achieved an average repair rate of 74.7\% across the 18 fault benchmarks, reaching up to 95.2\% in the best case.
In contrast, applying the original \arachne to the FFNs of ViT resulted in a much lower average repair rate of 17.1\%, with a maximum of 44.4\%.
\reptran also demonstrated competitive efficiency, with an average runtime of 476.19 seconds, which was faster than \arachne (620.93 seconds on average).
% A detailed analysis of hyperparameters of \reptran further revealed that the number of selected weights had a greater influence on repair effectiveness than the balance coefficient used in the fitness function.

The contributions of this study are as follows:
\begin{itemize}
    \item We propose \reptran, a novel DNN repair method tailored to Transformer models.
    \item We develop a weight selection strategy based on a \textit{neuron score} that quantifies behavioral differences in intermediate FFN neurons in the Transformer.
    % \item We are the first to apply \arachne, a prior DNN repair method for CNNs and RNNs, to Transformer models. We also introduce an enhanced baseline, \arachnew, which improves its weight selection capability.
    \item Our evaluation on 18 fault benchmarks demonstrates that \reptran outperforms \arachne and random baselines in both repair rate and execution time.
    \item All source code for our experiments is publicly available.\footnote{\url{https://github.com/posl/RepTran-replication}}
\end{itemize}

%% file: section/background.tex
\section{Background and Related Work} \label{sec:background}
In this section, we begin by introducing existing DNN repair methods.
Next, we describe the Transformer architecture to provide background for our work. 
Finally, we introduce existing work on internal analyses and editing of Transformer models that motivate our proposed repair method, \textsc{RepTran}.

\subsection{DNN Repair} \label{subsec:dnn_repair}
% Many studies in software engineering have explored methods for localizing and repairing defects in DNNs~\citemodelrepair.
The goal of DNN repair~\citemodelrepair is to achieve controlled modifications (\eg, fixing specific types of misclassifications).
% The motivation is to correct misbehaviors in trained models without full retraining, thereby saving time and computational resources.
Full retraining and repair serve complementary roles:
while full retraining pursues overall performance improvement when sufficient time and data are available, repair enables more controlled improvements even when such resources are limited.
% while full retraining is suitable for major updates when sufficient time and data are available, repair acts as a temporary patch when such resources are limited.

Among DNN repair methods, search-based approaches are common~\cite{sun2022ICSE,sohn2022tosem,tokui2022saner,calsi2023ICST,li2023GECCO}.
They typically involve two phases:
(1) a \textit{localization} phase that identifies the neurons or weights to be repaired, and
(2) a \textit{search} phase that uses metaheuristics to optimize parameter modifications.
% Arachne
For instance, \arachne~\cite{sohn2022tosem} considers both forward impact and gradient loss to identify faulty weights.
Forward impact measures the influence of the weight on the final output, while gradient loss measures its contribution to misbehaviors.
Faulty weights are identified as the Pareto front of these two metrics, and DE~\cite{storn1997de} is applied to modify their values.
% \arachne has been evaluated on CNNs for image classification and RNNs for text classification.
% CARE
Meanwhile, CARE~\cite{sun2022ICSE} identifies faulty neurons by measuring the causal effect of each neuron on the output.
The identified neurons are then refined using particle swarm optimization (PSO)~\cite{kennedy1995ICNN}.
% Their target tasks include backdoor removal, as well as fairness and safety improvements in FNNs and CNNs.
Our method, \textsc{RepTran}, follows the search-based approach but differs from existing methods in that it targets Transformer models and introduces a novel weight selection strategy for their FFNs.
% \textsc{RepTran} takes into account the unique characteristics of the FFNs in Transformers, which consist of two fully connected layers.
% We target FFNs in Transformers because they have the responsibility to process the output of the multi-head self-attention (MSA), and they account for two-thirds of the total parameters in the Transformer model~\cite{geva2021emnlp, wei2024nips}.

Another approach is to formulate the entire DNN repair process as an optimization problem (\eg, linear or quadratic programming) and then rely on a solver to produce repair results~\cite{sotoudeh2021PLDI, tao2023PLDI, sun2024tosem, nawas2024isaiv}.
For instance, PRDNN~\cite{sotoudeh2021PLDI} and APRNN~\cite{tao2023PLDI} translate the DNN repair problem into an linear programming problem, whereas AutoRIC~\cite{sun2024tosem} adopts a quadratic programming problem.
% The paper of AutoRIC reported a smaller accuracy drop compared to LP-based methods, and demonstrated its effectiveness in fairness and robustness repairs on FNNs and CNNs.
% PRoViT
More recently, PRoViT~\cite{nawas2024isaiv} was introduced as a provable repair method for ViT, ensuring correct classifications on a designated repair set.
PRoViT addresses the scalability issue of solver-based approaches by restricting repair targets to the final layer only.
It provides three variants that differ in how this final layer is modified:
(1) PRoViT$_\mathrm{LP}$ solves a linear program (LP) over the final layer;
(2) PRoViT$_\mathrm{FT}$ fine-tunes (FT) the final layer until it reaches 100\% accuracy on the repair set; and
(3) PRoViT$_\mathrm{FT+LP}$ first applies one epoch of fine-tuning to the final layer and then runs the above LP, so that the LP repairs whatever the single fine-tuning epoch has not yet fixed.

In our evaluation, we compare \textsc{RepTran} against representative methods from both categories: the search-based \arachne and the solver-based PRoViT.

\subsection{Transformer}
The Transformer~\cite{vaswani2017NIPS} consists of multi-head self-attention (MSA) and feed-forward networks (FFN), which are repeatedly stacked in multiple layers.
These mechanisms differ significantly from those in CNNs and RNNs, which have been the main focus of DNN repair thus far.

The FFN, which is the primary target of our repair method, consists of two fully connected layers with a non-linear activation function:
\begin{equation}
\text{FFN}(z) = W_{\text{aft}} (\sigma(W_{\text{bef}} z + b_{\text{bef}})) + b_{\text{aft}}, \label{eq:ffn}
\end{equation}
where $z \in \mathbb{R}^{D}$ denotes the input to the FFN (i.e., the output of the preceding MSA).
$W_{\text{bef}} \in \mathbb{R}^{D_{\text{hidden}} \times D}$ and $W_{\text{aft}} \in \mathbb{R}^{D \times D_{\text{hidden}}}$ are learnable weight matrices, while $b_{\text{bef}} \in \mathbb{R}^{D_{\text{hidden}}}$ and $b_{\text{aft}} \in \mathbb{R}^{D}$ are bias terms.
$\sigma$ represents a non-linear activation function such as ReLU or GELU.
Typically, $D_{\text{hidden}} = 4D$; that is, the FFN projects MSA outputs to a higher dimension via the first layer and then back to the original dimension via the second layer.
This structure is believed to contribute to the expressiveness of Transformer models~\cite{geva2021emnlp, dai2022ACL}.

\subsection{Internal Analysis and Editing of Transformer}

\textbf{Internal Analysis.} While Transformers have achieved remarkable success across various tasks, their internal mechanisms remain black boxes.
Therefore, a lot of studies have attempted to elucidate the functions of key components, such as MSA~\cite{michel2019nips,hao2021aaai,li2023tvcg} and FFN~\cite{geva2021emnlp,dai2022ACL}.
% This section introduces these existing studies emerging from the AI research community, yet they provide valuable insights for the development of repair methods for Transformer models.
This section introduces these existing studies; although they emerge from the AI research community, they still provide valuable insights for the development of repair methods for Transformer models.

% self-attention
Michel~\et~\cite{michel2019nips} have observed that a large percentage of attention heads can be removed at test time without significantly decreasing performance.
% They also introduced greedy algorithms for model pruning by defining a head importance score.
Similarly, Li~\et~\cite{li2023tvcg}~proposed pruning-based head importance metrics that assess the impact of attention heads both within the attention layers and on the final output for ViTs.
% Additionally, they developed a visualization tool to monitor MSA patterns in ViTs using these scores.

% feed-forward
Geva~\et~\cite{geva2021emnlp} have shown that FFNs in Transformer models operate as key-value memories.
% : in Eq.~\ref{eq:ffn}, the first fully connected layer with $W_{\text{bef}}$ corresponds to the key operation, whereas the second with $W_{\text{aft}}$ corresponds to the value operation.
% knowledge neurons
Building on this, Dai~\et~\cite{dai2022ACL} have shown that the intermediate neurons of FFNs (of dimension $D_{\text{hidden}}$) store learned knowledge, coining the term \textit{knowledge neurons}, with further studies~\cite{tang2024ACL, niu2024ICLR} extending this line of research.
These findings indicate that FFNs play a crucial role in the Transformer and offer insights relevant to DNN repair.
From a repair perspective, identifying knowledge neurons resembles a localization phase.
However, knowledge neurons are defined with respect to factual associations rather than the classification misbehaviors we aim to repair; moreover, while knowledge neurons are benign, DNN repair focuses on identifying problematic neurons or weights to be modified.
Motivated by these observations, we propose a novel repair method that targets FFNs of Transformer models.

% model editing (MEND, ROME)
\textbf{Model Editing.}
A related line of work edits Transformer models post hoc to change a specified factual association while preserving unrelated behavior~\cite{mitchell2021iclr,meng2022nips}.
Both methods take a single edit pair, i.e., a prompt and its new desired output, and write exactly that association into the model.
MEND~\cite{mitchell2021iclr} transforms the fine-tuning gradient of the edit pair into a weight update using auxiliary editor networks, whereas ROME~\cite{meng2022nips} directly writes a single key-value association into an FFN located by causal tracing, via a rank-one weight update.
In contrast, a repair request is not an edit pair with a known target association but a set of misclassified inputs sharing a misbehavior type, and the patch must generalize to unseen inputs of the same type rather than to paraphrases of one edited prompt.
Moreover, ROME presupposes the subject-relation-object structure of factual prompts and MEND requires editor training on an edit dataset, neither of which applies to our setting; the evaluation protocols also differ (paraphrase generalization vs. repair and break rates on held-out inputs).
% This goal is distinct from the repair setting studied in this paper.
% In model editing, the edit request is typically an individual factual association, such as a prompt and its desired answer, and generalization is evaluated on paraphrases.
% In contrast, \textsc{RepTran} treats a repair request as a set of misclassified inputs and reduces the corresponding misclassification on held-out test inputs while avoiding breaks on originally correct inputs.
% Thus, the two settings share the broader goal of post-hoc behavior modification, but differ in the modality (language vs. vision), the target of modification (an individual factual association vs. a set of misclassified inputs), and the evaluation protocol (paraphrase generalization vs. repair and break rates on held-out test inputs).
This distinction also explains our choice of baseline: PRoViT is a direct baseline because it targets ViT repair under the same repair and break evaluation, whereas adapting MEND and ROME would require redefining edit pairs and is left as future work.

%% file: section/method.tex
\input{image/reptran.tex}

\section{\textsc{RepTran}: Search-Based Repair of Transformer Models} \label{sec:method}

\figref{fig:reptran} illustrates the overall process of \textsc{RepTran}, our search-based repair method for Transformer models.
The input is the \textit{original Transformer model}, and the output is the \textit{patched Transformer model}.
\textsc{RepTran} consists of two main phases:
(1) a \textit{selection} phase, where we compute neuron- and weight-level scores in the FFN to identify suspicious weights; and 
% \footnote{Prior work often uses the term \textit{localization} phase, which is closely tied to fault localization for source code. Since we cannot obtain ground truth for faulty weights, our goal is not to pinpoint and evaluate faulty weights but to obtain a subset of weights that improves the subsequent repair. We therefore use the term \textit{selection} phase.}
(2) a \textit{search} phase, where a search algorithm modifies these weights, guided by a fitness function.
% We target the FFN because it is a functionally crucial component of the Transformer~\cite{geva2021emnlp,dai2022ACL}.
% This also reduces the number of target parameters to approximately two-thirds~\cite{geva2021emnlp}.

% \textbf{Our Technical Contributions.}
% % These phases are similar to those in existing search-based repair approaches~\cite{sun2022ICSE,sohn2022tosem,tokui2022saner,qi2023tosem,calsi2023ICST,li2023GECCO}, but our main contribution lies in the selection phase.
% Our main contributions lie in the selection phase.
% First, we focus on the FFN, which is a functionally crucial component of the Transformer~\cite{geva2021emnlp,dai2022ACL,kobayashi2020EMNLP}.
% The FFN consists of two linear transformations (\ie, $W_{\text{bef}}$ and $W_{\text{aft}}$) that first expand and then reduce the dimensionality, as shown in Eq.~\ref{eq:ffn}.
% This makes it not just a plain fully connected layer but a distinctive component of Transformers.
% This choice also reduces the number of target parameters to approximately two-thirds before the selection phase~\cite{geva2021emnlp}.
% Second, we propose an FFN-specific \textit{neuron score} to refine the weight selection.
% The neuron score is combined with the weight-level \textit{bidirectional score}~\cite{sohn2022tosem} to realize FFN-aware suspicious weight selection.
% We consider neuron-level selection too coarse-grained for precise repair because modifying an entire neuron affects all its associated weights simultaneously.

\subsection{Selection Phase} \label{sec:selection}
The goal of the selection phase is to narrow down the weight parameter space by identifying suspicious weights.
We target both $W_{\text{bef}}$ and $W_{\text{aft}}$ in the FFN (see Eq.~\ref{eq:ffn}) for selection.
Such feed-forward layers also appear in other architectures such as CNNs and RNNs, and our neuron score could in principle be computed on them as well.
We nonetheless focus on the FFNs of Transformers, whose intermediate neurons act as key-value memories that store semantic associations~\cite{geva2021emnlp, dai2022ACL}.
Our neuron score is inspired by this property, which motivates the positioning of \textsc{RepTran} as a Transformer-oriented repair method.
Extending \textsc{RepTran} to general DNN repair beyond Transformers is technically feasible but left for future work.

For each weight in these matrices, \textsc{RepTran} computes a \textit{weight suspiciousness score}, which combines two components: \textit{neuron score} and \textit{bidirectional score}.
\figref{fig:susptree} shows the composition of the weight suspiciousness score.
The neuron score (see \secref{sec:neuron_score}) quantifies the contribution of intermediate FFN neurons by using the mean and variance of their activations.
The bidirectional score (see \secref{sec:bi_score}) measures the forward impact and gradient loss of each weight.
The weight suspiciousness score is obtained by modulating its bidirectional score with the corresponding neuron score (see \secref{sec:weight_score}).
The bidirectional score alone does not account for the importance of individual neurons in the FFN of Transformer models; therefore, we explicitly incorporate this aspect into our approach.
Finally, this phase outputs suspicious weights, \ie, those with high suspiciousness scores.

\input{image/susptree.tex}

\subsubsection{Neuron Score} \label{sec:neuron_score}
As shown at the bottom of \figref{fig:susptree}, the neuron score quantifies the contribution of intermediate FFN neurons to the misbehavior.
Unlike knowledge neurons~\cite{dai2022ACL}, which rely on integrated gradients~\cite{sundararajan2017icml} requiring multiple forward and backward passes per input, our score is designed specifically for efficient repair by requiring only a single forward pass per input.
It consists of two components, \textit{VDiff} and \textit{MisAct}, described below.

\textit{VDiff} quantifies how differently each intermediate FFN neuron $h_{\text{mid}}^i$ behaves between correct and incorrect inputs, using the variance of activations.
Motivated by the use of variance as a coverage metric for DNNs~\cite{yuan2023ICSE} and as a risk indicator for model outputs~\cite{yuheng2025tse}, we define VDiff as follows:
\begin{equation*}
    \text{VDiff}(h_{\text{mid}}^i) = \left| \sum_{j=1}^{D_{\text{hidden}}} |\Sigma_{ij}^{\text{cor}}| - \sum_{j=1}^{D_{\text{hidden}}} |\Sigma_{ij}^{\text{mis}}| \right|,
\end{equation*}
where $\Sigma^{\text{cor}}$ and $\Sigma^{\text{mis}}$ are the covariance matrices of the hidden state $h_{\text{mid}} \in \mathbb{R}^{D_\text{hidden}}$, computed from the correctly classified samples ($\mathcal{I}_{\text{cor}}$) and the misclassified samples ($\mathcal{I}_{\text{mis}}$), respectively.
Each summation $\sum_j |\Sigma_{ij}|$ aggregates the neuron's own variance and absolute covariances with all other neurons, capturing the diversity of its behavior under a given input set.
VDiff thus identifies neurons whose behavioral diversity differs between correct and incorrect inputs.

% \footnotetext{Here, $h_{\text{mid}}^i$ is a neuron of the hidden state vector $h_{\text{mid}}$ in the FFN, \ie, $h_{\text{mid}} = (h_{\text{mid}}^1, h_{\text{mid}}^2, \cdots, h_{\text{mid}}^{D_{\text{hidden}}}) \in \mathbb{R}^{D_\text{hidden}}$.}

\textit{MisAct} measures the mean activation of neuron $h_{\text{mid}}^i$ under the misbehavior:
\begin{equation*}
    \text{MisAct}(h_{\text{mid}}^i) = \frac{1}{|\mathcal{I}_{\text{mis}}|}\sum_{(x, y) \in \mathcal{I}_{\text{mis}}} a_i(x),
\end{equation*}
where $a_i(x)$ is the activation value of neuron $h_{\text{mid}}^i$ for input $x$.
While VDiff captures variability, MisAct reflects how strongly the neuron fires on average under misbehavior.

We define the neuron score for $h_{\text{mid}}^i$ as follows:
% the product of the normalized VDiff and MisAct:

\begin{align*}
    \text{NeuronScore}(h_{\text{mid}}^i) &= \text{VDiff}_\text{norm}(h_{\text{mid}}^i) \times \text{MisAct}_\text{norm}(h_{\text{mid}}^i). \label{eq:neuronscore}
\end{align*}
Because VDiff and MisAct have different scales, we normalize each of them to $[0,1]$ (\ie, $\text{VDiff}_\text{norm}$ and $\text{MisAct}_\text{norm}$), ensuring that the neuron score lies in $[0,1]$.
We use the product of VDiff and MisAct to capture an \textit{and condition}: we emphasize neurons that show both a significant variance-based shift \emph{and} strong activation under misbehavior.
% We regard neurons with high neuron scores as suspicious and select them for the next step, weight selection.
% The number of selected neurons is hyperparameter, denoted as $N_n$.
This neuron score is combined with the bidirectional score, which is introduced next, to guide the selection of suspicious weights.

\subsubsection{Bidirectional Score} \label{sec:bi_score}
As shown in the gray parts in the middle of \figref{fig:susptree}, our selection phase uses the two existing metrics from \arachne~\cite{sohn2022tosem}: \textit{forward impact (FI)}, which measures the influence of a weight on the final output, and \textit{gradient loss (GL)}, which captures the gradient of the loss with respect to the weight.
% We refer readers to the original paper~\cite{sohn2022tosem} for detailed definitions.

\subsubsection{Weight Suspiciousness Score} \label{sec:weight_score}
As shown at the top of \figref{fig:susptree}, we compute the \textit{weight suspiciousness score} by combining the neuron score and the bidirectional score.
For each weight $w \in W_\text{bef} \cup W_\text{aft}$, there exists exactly one intermediate neuron $h_{\text{mid}}^i$ to which $w$ is connected (see the middle of \figref{fig:reptran}).
We modulate the bidirectional scores as follows:
\begin{align*}
\text{ModFI}(w) &= \text{FI}(w) \times \text{NeuronScore}(h_{\text{mid}}^i), \\
\text{ModGL}(w) &= \text{GL}(w) \times \text{NeuronScore}(h_{\text{mid}}^i).
\end{align*}
These modulated scores are normalized to $[0,1]$, and the final weight suspiciousness score is:
\begin{equation}
\text{WeightSusp}(w) = p \text{ModFI}(w) + (1-p) \text{ModGL}(w), \label{eq:weightsuspscore}
\end{equation}
where $p$ is a hyperparameter that controls the trade-off between the two scores.
We select the top-$N_w$ weights with the highest scores as suspicious weights.

\subsection{Search Phase}
After selecting suspicious weights, we apply a metaheuristic to repair the misbehavior without harming correct behavior.

% \subsubsection{Differential Evolution}
By default, \textsc{RepTran} employs DE~\cite{storn1997de}, although other search algorithms (\eg, PSO) can also be used.
DE maintains a population of candidate solutions, each representing a set of weight values for the selected weights.
These candidates are iteratively evolved through mutation and selection over multiple generations, guided by a fitness function.
Each candidate is initialized by sampling from a normal distribution whose mean and variance match those of the corresponding weight matrix ($W_\text{bef}$ or $W_\text{aft}$).
The search terminates after $G_{\text{max}}$ generations or when the best fitness stagnates for $G_{\text{stag}}$ generations.

% The fitness function balances two objectives: maintaining correct behavior on $\mathcal{I}_{\text{cor}}$ and reducing misbehavior on $\mathcal{I}_{\text{mis}}$:
% \begin{align}
%     \text{Fitness}(u) &= \frac{1}{|\mathcal{I}_{\text{cor}}|}\sum_{(x, y) \in \mathcal{I}_{\text{cor}}} \text{Score}_{u}(x, y) + \alpha \cdot \frac{1}{|\mathcal{I}_{\text{mis}}|}\sum_{(x, y) \in \mathcal{I}_{\text{mis}}} \text{Score}_{u}(x, y), \label{eq:fitness}
% \end{align}
% where $u$ is a candidate solution, $\alpha$ controls the trade-off between the two objectives, and $\text{Score}_{u}(x, y)$ returns 1 if the model with modified weights $u$ correctly predicts the label for input $x$, and $\frac{1}{1+\mathcal{L}(x, y)}$ otherwise.

% Fitnessの説明
% \subsubsection{Fitness Function} \label{sec:fitness}
The fitness function that DE aims to maximize is designed to reduce misbehavior while preserving correct behavior:
\begin{align}
    \text{Fitness}(u) &= \sum_{(x, y) \in \mathcal{I}_{\text{cor}}} \frac{S_{u}(x, y)}{|\mathcal{I}_{\text{cor}}|} + \alpha \sum_{(x, y) \in \mathcal{I}_{\text{mis}}} \frac{S_{u}(x, y)}{|\mathcal{I}_{\text{mis}}|}, \label{eq:fitness}
\end{align}
where $u$ is a candidate solution, $\alpha$ is a hyperparameter that balances maintaining correctness and reducing misbehavior, and $S_{u}(x, y)$ returns 1 if the model with modified weights $u$ correctly predicts the label for input $x$, and $\frac{1}{1+\mathcal{L}(x, y)}$ otherwise.
% This fitness function is similar to the one used in \arachne~\cite{sohn2022tosem}, except that we divide the score by the number of samples to account for the difference in sample size between $\mathcal{I}_{\text{cor}}$ and $\mathcal{I}_{\text{mis}}$.
Since correct predictions receive the maximum $S_{u}(x, y)$ of 1 while incorrect ones receive a value less than 1, maximizing this function drives DE toward solutions that preserve correct behavior while reducing misbehavior.

\begin{comment}
どっかで使うかも
% あるレイヤのFFNのパラメータ数は8D^2である．ViT-baseの場合はD=768であるため，パラメータ数は8D^2=4,718,592となる．
% 一方で，複数のDNN修正手法を比較した既存研究においては，対象モデルのパラメータ数は最大でも3,274,634である（CNN）．

残り書きたいこと：
一回ニューロンを絞り込んでから，重みを絞り込む理由は？
これらの手法の計算時間は？
\end{comment}

%% file: image/reptran.tex
\begin{figure}[t]
    \begin{center}
      \includegraphics[width=.9\linewidth]{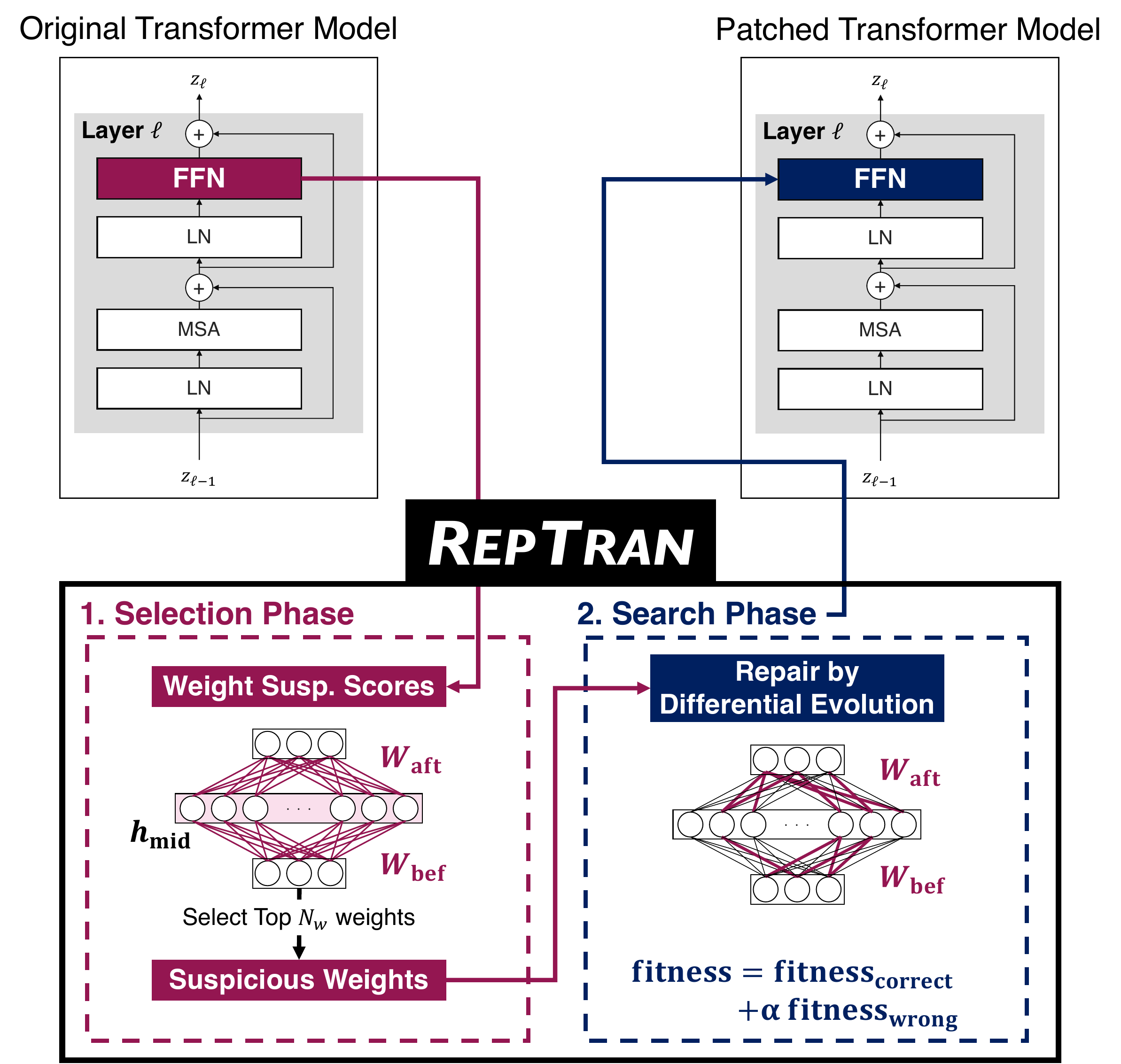}
      \caption{An overview of \textsc{RepTran}.}
      \label{fig:reptran}
    \end{center}
  \end{figure}

%% file: image/susptree.tex
\begin{figure}[t]
    \begin{center}
      \includegraphics[width=.9\linewidth]{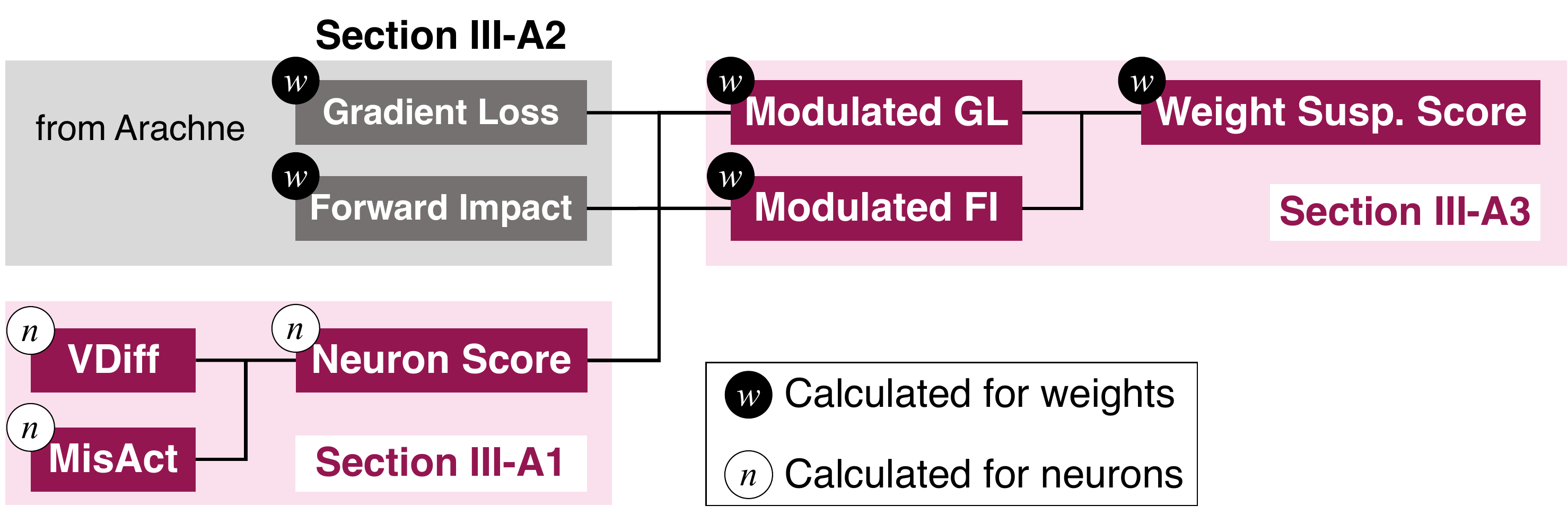}
      \caption{Composition of the weight suspiciousness score.}
      \label{fig:susptree}
    \end{center}
  \end{figure}

%% file: section/exp_setup.tex
\section{Experimental Design} \label{sec:exp_setup}

\subsection{Research Questions}
\label{sec:rqs}
% RQone, two, threeのコマンドを定義
\newcommand{\RQone}{\textbf{RQ1 (Effectiveness)}: How effective is \textsc{RepTran} compared with the search-based baselines?}
\newcommand{\RQtwo}{\textbf{RQ2 (Efficiency)}: How efficient is \textsc{RepTran} compared with the search-based baselines?}
\newcommand{\RQthree}{\textbf{RQ3 (Number of Selected Weights)}: How does the number of selected weights affect the repair performance?}
\newcommand{\RQfour}{\textbf{RQ4 (Balance Coefficients)}: How do the balance coefficients in each phase affect repair performance?}
\newcommand{\RQfive}{\textbf{RQ5 (Comparison with a ViT-specific method)}: How does \textsc{RepTran} compare with PRoViT?}

We formulate five RQs to evaluate the effectiveness and efficiency of \textsc{RepTran} in repairing DNNs, to investigate the impact of key parameters on its performance, and to compare it with a repair method tailored to ViTs.

\noindent
\RQone\\
\RQtwo\\
\RQthree\\
\RQfour\\
\textbf{RQ5 (Comparison with a ViT-specific method)}: How does \textsc{RepTran} compare with PRoViT?

\subsection{Datasets} \label{sec:exp_dataset}
We use two datasets in our experiments: \textit{CIFAR-100 (C100)}~\cite{krizhevsky2009learning} and \textit{Tiny-ImageNet (TinyImg)}~\cite{le2015tImg}, which are widely used for image classification tasks.
C100 is an image classification dataset with 100 classes, consisting of 50,000 training and 10,000 test samples of 32x32 color images.
TinyImg is a smaller version of ImageNet with 200 classes, consisting of 100,000 training, 10,000 validation, and 10,000 test samples of 64x64 color images.

\begin{comment} 今使ってないdatsets
    % \noindent \textbf{CIFAR-100-C}~\cite{hendrycks2019ICLR}.
    % CIFAR-100-C is a corrupted version of the CIFAR-100 dataset.
    % It consists of 15 types of common corruptions, such as noise, blur, and weather.
    % These corruptions are applied to the original 10,000 test samples of the CIFAR-100 dataset.
    % Each corruption type is applied at five different severity levels, resulting in a total of 750,000 corrupted images (10,000 images $\times$ 15 corruptions $\times$ 5 severity levels).
    % \noindent \textbf{Tiny-ImageNet-C}.
    % Tiny-ImageNet-C is a corrupted version of the Tiny-ImageNet dataset, which is a 200-class image classification dataset derived from ImageNet. 
    % Each sample is a 64x64 color image.
    % \todo[Fill this area after experiment for tiny-imagenet-c]
\end{comment}

To ensure consistency in data splitting, we follow prior studies on DNN repair~\cite{sohn2022tosem, ishimoto2025tosem} and adopt a three-way split, \ie, \textit{train}, \textit{repair}, and \textit{test}.
The train set is used to fine-tune the model, the repair set to identify and correct misbehavior, and the test set to evaluate how much the targeted misclassification has been reduced on data unseen during repair.
For both datasets, we split the original training set into train and repair sets with an 80:20 ratio.
For C100, the original test set is used as-is, resulting in 40,000/10,000/10,000 samples for train/repair/test.
For TinyImg, since the original test set lacks ground-truth labels,
% \footnote{The ground-truth labels of the test set are not publicly released, as they are reserved for leaderboard evaluation.
we use the original validation set as the test set, resulting in 80,000/20,000/10,000 samples.

\subsection{Transformer Models} \label{sec:tf_models}
We use ViT~\cite{dosovitskiy2021ICLR} as the subject model, a representative Transformer architecture widely adopted in vision tasks.
We use the publicly available weights on Hugging Face.\footnote{\url{https://huggingface.co/google/vit-base-patch16-224}}
The model consists of 12 Transformer encoder blocks, each containing a FFN with a hidden dimension of 3,072 (\ie, $D_{\text{hidden}} = 3072, D = 768$).
We focus on ViT because vision tasks often involve safety-critical applications such as autonomous driving~\cite{hu2023cvpr} and medical imaging~\cite{manzari2023cbm}, and an industry report~\cite{nakagawa2023saner} emphasizes the demand for practical DNN repair methods in this domain.
Note that \textsc{RepTran} can be applied, in principle, to any Transformer architecture that employs FFNs.

In the experiment, we fine-tune the ViT for two epochs.
%  by adding a classification head specific to the dataset.
After fine-tuning, the models achieved test accuracies of 91.18\% on C100 and 86.34\% on TinyImg.
These models serve as the original models to be repaired.
% Considering that models are expected to be well-trained before deployment, our setup reflects such real-world repair scenarios.
% In contrast, existing studies~\cite{qi2023tosem, hu2025tosem} used CNN-based models that were underfitting on these datasets, with test accuracies around 40\% on TinyImg and 60--70\% on C100.
Our setup reflects more realistic scenarios in which even well-trained models may still need to be repaired for specific types of misbehavior.

\input{table/fault_bench_size.tex}

\subsection{Fault Benchmarks} \label{sec:fault_type}
% We describe how to create fault benchmarks from the datasets.
A fault benchmark refers to a set of misclassified samples that are targeted for repair.
% Note that it is defined differently depending on a target misclassification.
The target misclassification is identified by specifying its \textit{type} and \textit{rank}.

First, we define three types of misclassification based on the pair of predicted and true labels $(c_\text{pred}, c_\text{true})$ as follows:
\begin{enumerate}
    \item \textit{SRC-TGT.} Misclassification where the model predicts a specific class $src$ for inputs whose true label is another specific class $tgt$:
    \begin{equation*}
        (c_\text{pred}, c_\text{true}) = (src, tgt), \quad \text{where } src \neq tgt.
    \end{equation*}
    % For example, all cases where images of cats ($tgt$) are misclassified as dogs ($src$).

    \item \textit{TGT-FP} (False positives for $tgt$). Misclassification where the model wrongly predicts $tgt$ for inputs not in $tgt$:
    \begin{equation*}
        \exists y \neq tgt,\quad (c_\text{pred}, c_\text{true}) = (tgt, y).
    \end{equation*}
    % For example, all cases where the model predicts cat ($tgt$) for images that are actually not cat.

    \item \textit{TGT-FN} (False negatives for $tgt$). Misclassification where the model fails to predict $tgt$ for inputs labeled as $tgt$:
    \begin{equation*}
        \exists y' \neq tgt,\quad (c_\text{pred}, c_\text{true}) = (y', tgt).
    \end{equation*}
    % For example, all cases where images labeled as cat ($tgt$) are misclassified into other classes.

\end{enumerate}
We consider these benchmark types to reflect real-world scenarios where developers are particularly concerned about classes that produce a lot of false positives or false negatives.

Next, we define the rank of misclassification for each type of misclassification.
For the \textit{SRC-TGT}, we count the number of misclassified samples for each pair of (\textit{src}, \textit{tgt}) and rank these pairs in descending order of frequency.
This allows us to identify the most frequent misclassification patterns between specific class pairs.
For the \textit{TGT-FP} and \textit{TGT-FN}, we compute the precision and recall for each class, respectively, and rank the classes in ascending order.
% This reflects real-world scenarios where developers are particularly concerned about classes that produce many false positives or false negatives.
In our experiments, we target the top-3 ranks (\ie, rank 1, 2, and 3).
By specifying the misclassification type and its rank, we can identify a small set of target samples that represent the target misclassification.
This set corresponds to $\mathcal{I}_{\text{mis}}$ in \secref{sec:neuron_score}, while the set of correctly classified samples corresponds to $\mathcal{I}_{\text{cor}}$.

\tabref{tab:fault_bench_size} shows the number of misclassified samples used in this study.
We obtain a total of 18 fault benchmarks (2 datasets $\times$ 3 types $\times$ 3 ranks).
While the misclassification types and ranks are identified based on the repair set, the generalization performance of the repair is evaluated on a disjoint test set.
Therefore, for each type and rank of misclassification in the repair set (\ie, $|\mathcal{I}_\text{mis}^{\text{repair}}|$), we also report the number of corresponding misclassified samples in the test set (\ie, $|\mathcal{I}_\text{mis}^{\text{test}}|$).
% Note that for the \textit{TGT-FP} and \textit{TGT-FN} types, the ranks are determined based on proportion-based scores such as precision and recall, rather than raw counts.
% As a result, the ranks do not necessarily correspond to the descending order of the number of misclassified samples.

\begin{comment} Icorからのサンプリングの話
    % Meanwhile, $\mathcal{I}_{\text{cor}}$ consists of a subset of correctly classified samples.
    % Using all correctly classified samples would be computationally expensive, particularly for fitness calculations described in Section~\ref{sec:fitness}.
    % To improve efficiency, we randomly select 200 samples from the correctly classified samples and use them as $\mathcal{I}_{\text{cor}}$.
    % This sample size is empirically determined, balancing experimental efficiency and representativeness.
    % Additionally, we ensure that at least one sample from each label is included during random sampling.
\end{comment}

\subsection{Baselines} \label{sec:baselines}
% The primary novelty of our method lies in the selection phase (see \secref{sec:selection}).
% To highlight its effectiveness, we compare our approach against several baselines that employ the same search phase but differ in how they select weights.

The first baseline is a \textit{random} strategy, where a fixed number of weights are randomly chosen from the set $W_{\text{bef}} \cup W_{\text{aft}}$.

The second baseline is \arachne~\cite{sohn2022tosem}, a state-of-the-art DNN repair method that selects weights on the Pareto front defined by the forward impact and the gradient loss (gray blocks in \figref{fig:susptree}).
Although the original \arachne does not support Transformer models, we apply it to the FFN layers, which can be regarded as feed-forward networks composed of two linear transformations (\ie, $W_{\text{bef}}$ and $W_{\text{aft}}$).

The third baseline is PRoViT~\cite{nawas2024isaiv}, a repair method designed specifically for ViTs, which we use in RQ5.
We evaluate its three variants (PRoViT$_\mathrm{LP}$, PRoViT$_\mathrm{FT}$, and PRoViT$_\mathrm{FT+LP}$), all targeting the same final-block FFN as \textsc{RepTran} so that the methods differ only in how they modify that module.

\subsection{Configuration of Repair Methods} \label{sec:exp_config}
% \tabref{tab:exp_config} summarizes the configurations for repair.
% Since the baselines differ only in the selection phase, we used a common configuration for the search phase across all baselines.

\textbf{Selection Phase.}
The parameter $p$ in Eq.~\ref{eq:weightsuspscore}, which balances the contributions of ModFI and ModGL in the suspiciousness score, is set to 0.5 for RQ1--3.
This means that the ModFI and ModGL are considered equally important.
We investigate the sensitivity of this choice in RQ4 ($p \in \{0.1, 0.5, 0.9\}$).
% , which aligns with the design of \arachne, where both FI and GL are taken into account via the Pareto front.

The number of selected weights $N_w$ is set to one of the values in \{11, 236, 472, 944\}.
Here, $N_w = 11$ corresponds to the average number of weights selected by \arachne.
In contrast, $N_w = 236, 472, 944$ correspond to 0.005\%, 0.01\%, and 0.02\% of the total number of weight parameters in a FFN layer, respectively, and are used in RQ3 to analyze the impact of the number of selected weights.

We target only the final layer of the Transformer encoder block (\ie, $L=12$) for repair.
Although our method can, in principle, be extended to other layers or even multiple layers, we adopt this design for the following reasons.
First, we follow prior work such as PRoViT~\cite{nawas2024isaiv}, which also focuses on the final layer to reduce the number of target parameters.
Second, a recent study~\cite{dai2022ACL} revealed that most of the knowledge neurons are concentrated in the final layer, suggesting the importance of this layer.
Therefore, our choice is not only practically motivated but also supported by prior findings that highlight the semantic significance of the final layer.

Moreover, to empirically support this design choice, we conducted an experiment to examine which layers are most suitable for repair.
We computed the neuron scores across all layers.
\figref{fig:neuron_scores_grid} presents the layer-wise mean neuron scores for each misclassification type and rank on C100; TinyImg exhibited the same trend and is omitted for space, with the full results available in our replication package.
We observe that the final layer consistently shows significantly higher scores than the preceding layers across misclassification types and ranks.
This observation is similar to prior findings~\cite{dai2022ACL} that emphasize the concentration of the knowledge neurons in the final layer.
This suggests that the final layer captures the most distinguishable behavior between correctly and incorrectly classified inputs, making it a prime target for repair.

\input{image/neuron_scores_grid.tex}

\textbf{Search Phase.}
We use the same DE parameters as \arachne~\cite{sohn2022tosem}, including the population size $N_p$, scaling factor $F$, and crossover rate $CR$.
The parameters $G_{\text{max}}$ and $G_{\text{stag}}$, which control the termination conditions of the search phase, are empirically set to 50 and 10, respectively, to avoid excessive repair time.
The coefficient $\alpha$ in the fitness function (Eq.~\ref{eq:fitness}) balances the trade-off between maintaining correct behavior and reducing misbehavior.
We investigate the impact of $\alpha$ on the performance of the patched model in RQ4.
In the other RQs, we fix it to 10, following prior studies~\cite{le2011TSE,sohn2022tosem}.
To account for the randomness in the search phase introduced by DE, we repeat each experiment five times (\#runs=5) under the same configuration.

\subsection{Evaluation Metrics} \label{sec:eval_metrics}
We apply each repair method on the repair set and evaluate the patched model $M'$ on the test set, using three metrics: \textit{repair rate} ($RR$), \textit{break rate} ($BR$), and \textit{total execution time} ($T_{\text{tot}}$).
Let $\mathcal{I}^{\text{test}}_{\text{cor}}$ and $\mathcal{I}^{\text{test}}_{\text{mis}}$ denote the test-set samples correctly classified by the original model and the target misclassified samples, respectively.

The repair rate ($RR$) is the fraction of target misclassifications corrected in the test set:
\begin{equation*}
    RR = \frac{\left| \{ (x, y) \in \mathcal{I}^{\text{test}}_{\text{mis}} \mid M'(x) = y \} \right|}{\left| \mathcal{I}^{\text{test}}_{\text{mis}} \right|}.
\end{equation*}

The break rate ($BR$) captures the side effects of the repair, i.e., the fraction of originally correct samples that the patched model misclassifies:
\begin{equation*}
    BR = \frac{\left| \{ (x, y) \in \mathcal{I}^{\text{test}}_{\text{cor}} \mid M'(x) \ne y \} \right|}{\left| \mathcal{I}^{\text{test}}_{\text{cor}} \right|}.
\end{equation*}

The total repair time ($T_{\text{tot}}$) is the overall time for the repair process.
For \arachne and \textsc{RepTran}, it is the sum of the selection phase ($T_{\text{select}}$) and the search phase ($T_{\text{repair}}$); for the random baseline, $T_{\text{tot}}$ equals the search time.

% \subsection{Implementation} \label{sec:imp}
% All experiments were run on a machine equipped with an Intel Core i9‑14900K CPU and an NVIDIA GeForce RTX 4090 GPU.
% Detailed version information for all libraries is available in our replication package.
% We used Python 3.7.16 and loaded pretrained Transformer models using the HuggingFace Transformers library~\cite{wolf2019arXiv} (version: transformers 4.30.2).
% For implementing the repair methods, we used PyTorch~\cite{paszke2019nips} (version: torch 1.13.1) for handling model parameters and DEAP~\cite{fortin2012jmlr} (version: deap 1.3.3) for DE.
% % Our source code is publicly available.\footnote{\url{https://github.com/posl/RepTran-replication}}

%% file: table/fault_bench_size.tex
\begin{table}[t]
  \centering
  \small
  \caption{
    Number of misclassified samples for repair and test sets.
  }
  \label{tab:fault_bench_size}
  \resizebox{.8\columnwidth}{!}{
  \begin{tabular}{l|*{3}{c}|*{3}{c}|*{3}{c}}
    \toprule
    Type (Rank)
    & \multicolumn{3}{c|}{\textit{SRC-TGT}\,(1/2/3)} & \multicolumn{3}{c|}{\textit{TGT-FP}\,(1/2/3)} & \multicolumn{3}{c}{\textit{TGT-FN}\,(1/2/3)} \\
    \midrule
    \rowcolor{silver}
    \multicolumn{10}{c}{C100} \\
    \midrule
    $|\mathcal{I}_\text{mis}^{\text{repair}}|$ 
      & 19 & 16 & 14 
      & 27 & 32 & 31 
      & 27 & 33 & 27 \\
    $|\mathcal{I}_\text{mis}^{\text{test}}|$ 
      & 21 & 8 & 8 
      & 34 & 23 & 42 
      & 28 & 25 & 29 \\
    \midrule
    \rowcolor{silver}
    \multicolumn{10}{c}{TinyImg} \\
    \midrule
    $|\mathcal{I}_\text{mis}^{\text{repair}}|$ 
      & 33 & 20 & 18 
      & 56 & 46 & 41 
      & 39 & 32 & 34 \\
    $|\mathcal{I}_\text{mis}^{\text{test}}|$ 
      & 14 & 12 & 9 
      & 25 & 15 & 25 
      & 20 & 13 & 16 \\
    \bottomrule
  \end{tabular}
  }
\end{table}

%% file: image/neuron_scores_grid.tex
\begin{figure}[t]
  \centering
  \includegraphics[width=.9\linewidth]{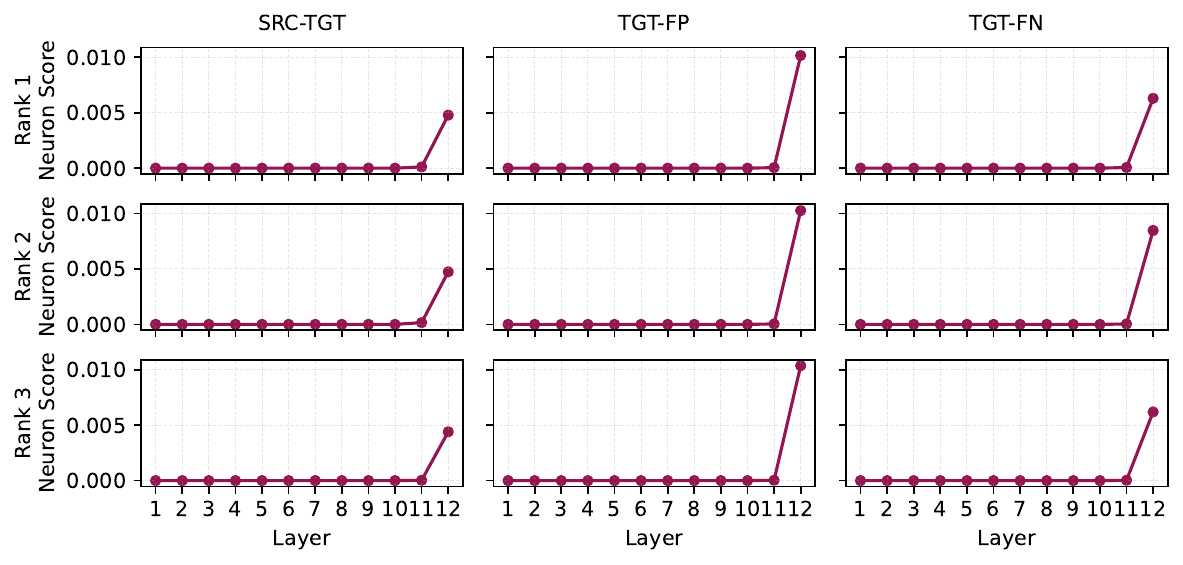}
  \caption{Layer-wise mean neuron scores for C100.}
  \label{fig:neuron_scores_grid}
\end{figure}

%% file: section/evaluation.tex
\section{Evaluation} \label{sec:evaluation}

All experiments were run on a machine equipped with an Intel Core i9‑14900K CPU and an NVIDIA GeForce RTX 4090 GPU.
Detailed version information for all libraries is available in our replication package.

\input{section/rqs/rq1.tex}
\input{section/rqs/rq2.tex}
\input{section/rqs/rq3.tex}
\input{section/rqs/rq4.tex}
\input{section/rqs/rq5.tex}

%% file: section/rqs/rq1.tex
\subsection{RQ1: Effectiveness of \textsc{RepTran}}

\textbf{Approach.}
We evaluate \textsc{RepTran} by comparing it with \arachne and random baselines.
\textsc{RepTran} selects 472 weights (0.01\% of FFN parameters), whereas \arachne selects 11 on average.
Since \arachne selects weights on the Pareto front, the number of selected weights cannot be controlled.
Experiments that control the number of selected weights are conducted in RQ3;
in RQ1, we configure each method to reflect settings under which it is expected to perform well.
We also prepare two random baselines:
Random$_R$, which selects the same number of weights as \textsc{RepTran} (472), and
Random$_A$, which selects the same number as \arachne (11),
to assess whether each method outperforms random selection under an equivalent weight budget.

\input{table/test_perf.tex}

\textbf{Results.}
\tabref{tab:test_perf} shows the repair generalization performance of each method on C100 and TinyImg.
Each cell in the table reports the average value over five runs.

\textbf{\textsc{RepTran} achieves the highest repair rate in all 18 cases.}
Its performance margin over the second-best method is substantial in most cases.
At the same time, \textsc{RepTran} consistently records the largest break rate.
This suggests that it tends to select weights that strongly influence model outputs.
Such weights are effective in correcting misclassifications, but they also increase the risk of breaking correct classifications.
Even so, the trade-off appears reasonable, given that the maximum break rate stays below 5.7\%, while the maximum repair rate reaches as high as 95.2\%.
% Random$_A$ stands in contrast to \textsc{RepTran}, recording the lowest repair rate and break rate in all 18 cases.
% Because it randomly selects only 11 weights, its effect on the model is minimal, resulting in very limited repair and very few side effects.
% Because it randomly selects only 11 weights, its effect on the model is minimal, resulting in a repair rate and break rate that are both close to zero.

\textbf{\arachne exhibits a more conservative behavior than \textsc{RepTran}.}
\arachne has a performance level between \textsc{RepTran} and the random baselines.
While its repair rate is substantially lower than that of \textsc{RepTran}, it still outperforms the random baselines.
In fact, \arachne ranks second in repair rate in 17 out of 18 cases.
% , with the only exception being the \textit{TGT-FP} case in TinyImg (rank 3), where it falls to third.
This result suggests that selecting weights based on the Pareto front may be too limited to achieve substantial repairs for the FFNs in Transformer models, though it is more effective than random selection.

\textbf{Both \textsc{RepTran} and \arachne exhibit higher repair and break rates compared to the random baselines, as seen in the comparisons between \textsc{RepTran} vs. Random$_R$ and \arachne vs. Random$_A$.}
This suggests that they can identify weights that are more likely to contribute to fixing misclassifications than those selected randomly, demonstrating the effectiveness of the selection phase.
At the same time, the result indicates the inherent difficulty of avoiding side effects when selecting such influential weights.

To statistically evaluate the differences in repair and break rates across repair methods, we performed the Wilcoxon signed-rank test~\cite{wilcoxon1992test}.
As shown in \tabref{tab:stat_test_perf}, we compared each pair of methods for every dataset, using 9 metric values per comparison (3 ranks $\times$ 3 misclassification types, where each value represents the average across 5 runs).
The table reports the statistical significance and the effect size, measured by Cliff's $\delta$~\cite{cliff1993dominance}.
It ranges from $-1$ to $+1$: positive values indicate that the first method in \tabref{tab:stat_test_perf} tends to have larger metric values than the second, whereas negative values indicate the opposite.
An effect size is considered large when $|\delta| > 0.474$~\cite{romano2006AFAI}.
All reported $p$-values are corrected using the Holm method~\cite{holm1979test} to control the family-wise error rate.

\textbf{\tabref{tab:stat_test_perf} shows that high repair and break rates of \textsc{RepTran} are statistically significant.}
In fact, all comparisons involving \textsc{RepTran} yield an effect size of 1.00 and $p < .05$.
This trade-off between repair and break rates is also observed in other comparisons.
% 唯一の例外は\arachne vs. Random$_A$, where \arachne achieves a higher repair rate but a lower break rate than Random$_A$ from the effect size.だけど統計的有意差はないケースが多い．
The only exception is found in the comparison between \arachne and Random$_R$, where \arachne shows a higher repair rate and a lower break rate based on the sign of the effect size;
however, the effect sizes are smaller than those involving \reptran, and the differences in break rate are not statistically significant.

\summaryblock{Answer to RQ1}{
\textsc{RepTran} outperforms all baselines in repair rate across all 18 cases, achieving up to 95.2\%.
Its superior effectiveness comes with a higher break rate, though the maximum remains below 5.7\%.
Statistical tests confirm that these differences are statistically significant in all comparisons involving \textsc{RepTran}.
}

\input{table/stat_test_perf.tex}

%% file: table/test_perf.tex
\begin{table}[t]
  \centering
  \caption{RQ1 - Repair generalization performance.
  Row-wise ranking is highlighted with Gold (1st), Silver (2nd), and Bronze (3rd). 
  Arrows $\uparrow$ ($\downarrow$) indicate that higher (lower) values are better.}
  \definecolor{gold}{HTML}{FFD700}
  \definecolor{bronze}{HTML}{B36B00}
  \definecolor{greycell}{HTML}{C0C0C0}
  \setlength{\tabcolsep}{4pt}
  \resizebox{\columnwidth}{!}{
  \begin{tabular}{ll*{4}{cc}}
  \toprule
  \multirow{2}{*}{Dataset} & \multirow{2}{*}{Rank / Type} &
  \multicolumn{2}{c}{\textsc{RepTran}} &
  \multicolumn{2}{c}{\arachne} &
  \multicolumn{2}{c}{Random$_R$} &
  \multicolumn{2}{c}{Random$_A$}\\
  \cmidrule(lr){3-4}\cmidrule(lr){5-6}\cmidrule(lr){7-8}\cmidrule(lr){9-10}
   &  & $RR \uparrow$ & $BR \downarrow$ & $RR \uparrow$ & $BR \downarrow$ & $RR \uparrow$ & $BR \downarrow$ & $RR \uparrow$ & $BR \downarrow$\\
  \midrule
  \multirow{9}{*}{C100}
   & 1 / \textit{SRC-TGT} & \cellcolor{gold}95.2\% & 0.9\% & \cellcolor{greycell}19.0\% & \cellcolor{greycell}0.2\% & \cellcolor{bronze}6.7\% & \cellcolor{greycell}0.2\% & 0.0\% & \cellcolor{gold}0.0\%\\
   & 1 / \textit{TGT-FP} & \cellcolor{gold}68.2\% & 1.4\% & \cellcolor{greycell}17.6\% & \cellcolor{greycell}0.2\% & \cellcolor{bronze}7.6\% & \cellcolor{greycell}0.2\% & 0.0\% & \cellcolor{gold}0.0\%\\
   & 1 / \textit{TGT-FN} & \cellcolor{gold}94.3\% & 1.3\% & \cellcolor{greycell}25.0\% & \cellcolor{greycell}0.2\% & \cellcolor{bronze}7.9\% & \cellcolor{greycell}0.2\% & 0.0\% & \cellcolor{gold}0.0\%\\ \cmidrule{2-10}
   & 2 / \textit{SRC-TGT} & \cellcolor{gold}80.0\% & 0.9\% & \cellcolor{greycell}37.5\% & \cellcolor{greycell}0.1\% & \cellcolor{bronze}12.5\% & \cellcolor{greycell}0.1\% & 7.5\% & \cellcolor{gold}0.0\%\\
   & 2 / \textit{TGT-FP} & \cellcolor{gold}52.2\% & 1.4\% & \cellcolor{greycell}13.0\% & \cellcolor{greycell}0.1\% & \cellcolor{bronze}1.7\% & \cellcolor{greycell}0.1\% & 0.0\% & \cellcolor{gold}0.0\%\\
   & 2 / \textit{TGT-FN} & \cellcolor{gold}92.8\% & 2.4\% & \cellcolor{greycell}20.0\% & \cellcolor{greycell}0.2\% & \cellcolor{bronze}16.8\% & \cellcolor{greycell}0.2\% & 0.0\% & \cellcolor{gold}0.0\%\\ \cmidrule{2-10}
   & 3 / \textit{SRC-TGT} & \cellcolor{gold}82.5\% & 0.9\% & \cellcolor{greycell}0.0\% & \cellcolor{greycell}0.1\% & \cellcolor{greycell}0.0\% & \cellcolor{greycell}0.1\% & 0.0\% & \cellcolor{gold}0.0\%\\
   & 3 / \textit{TGT-FP} & \cellcolor{gold}68.6\% & 1.3\% & \cellcolor{greycell}7.1\% & \cellcolor{gold}0.0\% & \cellcolor{bronze}4.3\% & \cellcolor{greycell}0.1\% & 0.0\% & \cellcolor{gold}0.0\%\\
   & 3 / \textit{TGT-FN} & \cellcolor{gold}86.9\% & 1.6\% & \cellcolor{greycell}6.9\% & \cellcolor{greycell}0.1\% & \cellcolor{greycell}6.9\% & \cellcolor{greycell}0.1\% & 0.0\% & \cellcolor{gold}0.0\%\\
  \midrule
  \multirow{9}{*}{TinyImg}
   & 1 / \textit{SRC-TGT} & \cellcolor{gold}91.4\% & 0.7\% & \cellcolor{greycell}28.6\% & \cellcolor{greycell}0.1\% & \cellcolor{bronze}21.4\% & \cellcolor{bronze}0.2\% & 0.0\% & \cellcolor{gold}0.0\%\\
   & 1 / \textit{TGT-FP} & \cellcolor{gold}14.4\% & 1.2\% & \cellcolor{greycell}9.6\% & \cellcolor{bronze}0.3\% & \cellcolor{bronze}4.0\% & \cellcolor{greycell}0.2\% & 0.0\% & \cellcolor{gold}0.0\%\\
   & 1 / \textit{TGT-FN} & \cellcolor{gold}78.0\% & 1.4\% & \cellcolor{greycell}20.0\% & \cellcolor{greycell}0.1\% & \cellcolor{bronze}16.0\% & \cellcolor{bronze}0.2\% & 1.0\% & \cellcolor{gold}0.0\%\\  \cmidrule{2-10}
   & 2 / \textit{SRC-TGT} & \cellcolor{gold}90.0\% & 0.7\% & \cellcolor{greycell}33.3\% & \cellcolor{greycell}0.1\% & \cellcolor{bronze}15.0\% & \cellcolor{bronze}0.2\% & 0.0\% & \cellcolor{gold}0.0\%\\
   & 2 / \textit{TGT-FP} & \cellcolor{gold}42.7\% & 1.1\% & \cellcolor{greycell}0.0\% & \cellcolor{greycell}0.1\% & \cellcolor{greycell}0.0\% & \cellcolor{bronze}0.2\% & 0.0\% & \cellcolor{gold}0.0\%\\
   & 2 / \textit{TGT-FN} & \cellcolor{gold}90.8\% & 5.7\% & \cellcolor{greycell}16.9\% & \cellcolor{bronze}0.4\% & \cellcolor{bronze}0.0\% & \cellcolor{greycell}0.2\% & 0.0\% & \cellcolor{gold}0.0\%\\ \cmidrule{2-10}
   & 3 / \textit{SRC-TGT} & \cellcolor{gold}86.7\% & 0.4\% & \cellcolor{greycell}44.4\% & \cellcolor{greycell}0.1\% & \cellcolor{bronze}20.0\% & \cellcolor{bronze}0.2\% & 0.0\% & \cellcolor{gold}0.0\%\\
   & 3 / \textit{TGT-FP} & \cellcolor{gold}39.2\% & 0.8\% & \cellcolor{bronze}2.4\% & \cellcolor{greycell}0.2\% & \cellcolor{greycell}5.6\% & \cellcolor{greycell}0.2\% & 0.0\% & \cellcolor{gold}0.0\%\\
   & 3 / \textit{TGT-FN} & \cellcolor{gold}90.0\% & 3.1\% & \cellcolor{greycell}6.2\% & \cellcolor{greycell}0.1\% & \cellcolor{bronze}3.8\% & \cellcolor{bronze}0.2\% & 0.0\% & \cellcolor{gold}0.0\%\\ \midrule
   Average & & 74.7\% & 1.5\% & 17.1\% & 0.2\% & 8.3\% & 0.2\% & 0.5\% & 0.0\%\\ 
   \bottomrule
  \end{tabular}
  }
  \label{tab:test_perf}
\end{table}

%% file: table/stat_test_perf.tex
\begin{table}[t]
\centering
\caption{
RQ1 - Wilcoxon signed-rank tests on $RR$ and $BR$.
Each cell reports Cliff's $\delta$.
Significance levels: $^{\ast}p{<}.05$, $^{\ast\ast}p{<}.01$.
Abbreviations: Rep = \reptran, Ara = \arachne, Rand$_R$ = Random$_R$, Rand$_A$ = Random$_A$.
}
\label{tab:stat_test_perf}
\resizebox{0.9\columnwidth}{!}{
    \begin{tabular}{lllllllll}
        \toprule
        Dataset & Metric & 
        \makecell{Rep\\vs.\ Ara} & \makecell{Rep\\vs.\ Rand$_R$} & \makecell{Rep\\vs.\ Rand$_A$} &
        \makecell{Ara\\vs.\ Rand$_R$} & \makecell{Ara\\vs.\ Rand$_A$} & \makecell{Rand$_R$\\vs.\ Rand$_A$} \\
        \midrule
        \multirow{2}{*}{C100}
        & $RR \uparrow$ & +1.00 * & +1.00 * & +1.00 * & +0.78 * & +0.89 * & +0.89 * \\
        & $BR \downarrow$ & +1.00 * & +1.00 * & +1.00 * & -0.33   & +1.00 * & +1.00 * \\
        \midrule
        \multirow{2}{*}{TinyImg}
        & $RR \uparrow$ & +1.00 * & +1.00 * & +1.00 * & +0.67 * & +0.89 * & +0.78 * \\
        & $BR \downarrow$ & +1.00 * & +1.00 * & +1.00 * & -0.56   & +1.00 * & +1.00 * \\
        \bottomrule
    \end{tabular}
}
\end{table}

%% file: section/rqs/rq2.tex
\subsection{RQ2: Efficiency of \textsc{RepTran}}

\textbf{Approach.}
We use the same patched models as in RQ1 but focus on the total repair time instead of repair and break rates.
For \textsc{RepTran} and \arachne, we report the total repair time as well as the time spent in the selection and search phases.
For the random baselines, only the total repair time is reported.

\textbf{Results.}
\tabref{tab:exec_time} shows the total repair time of \textsc{RepTran}, \arachne, Random$_R$, and Random$_A$ on C100 and TinyImg.
Each cell in the table reports the average value over five runs.
Bold numbers in the \(T_{\text{select}}\) and \(T_{\text{repair}}\) columns mark the faster method between \textsc{RepTran} and \arachne for each case.
We also show the statistical significance and the effect size of the differences in the total repair time across methods in \tabref{tab:stat_exec_time}.

\input{table/exe_time.tex}
\input{table/stat_exe_time.tex}

\textbf{From \tabref{tab:exec_time}, \textsc{RepTran} is consistently faster than \arachne across all 18 cases.}
When examined by phase, \textsc{RepTran} is faster than \arachne in the search phase in all cases, and also in the selection phase in 17 out of 18 cases.
Wilcoxon test results in \tabref{tab:stat_exec_time} confirm this advantage: 
for both datasets, the RepTran\,vs.\,\arachne comparison yields an effect size of $-1.00$ with statistical significance ($p{<}.05$).

This efficiency is notable given that the number of selected weights, which is the number of variables in the DE algorithm, is larger for \textsc{RepTran} than for \arachne (472\,vs.\,11).
This is not a trivial outcome, as shown by the comparison between Random$_R$ and Random$_A$ in the rightmost column of \tabref{tab:stat_exec_time}, where Random$_R$, which uses more variables, is significantly slower.
These results suggest that \textsc{RepTran} effectively identifies weights for faster convergence, enabling early stopping and shorter repair time despite handling more variables.

To further examine the efficiency of each method, we analyze the number of steps for the search phase, as shown in \figref{fig:num_steps}.
\textbf{\textsc{RepTran} terminated early in 87.8\% of runs, accounting for its short repair time.}
Similarly, Random$_A$, which also exhibited short repair time, triggered early stopping in 91.1\% of the runs.
However, the reasons for early stopping differ between \textsc{RepTran} and Random$_A$, especially in light of the results from RQ1.
For \textsc{RepTran}, early termination is due to rapid convergence to high fitness values, which aligns with its strong repair performance observed in RQ1.
By contrast, Random$_A$ stops early because the selected weights have minimal effect on models, resulting in little to no improvement in fitness and thus early termination for a negative reason.

\summaryblock{Answer to RQ2}{
\textsc{RepTran} is faster than \arachne in all 18 cases, despite optimizing more variables (472 vs. 11).
It achieves early stopping in 87.8\% of runs, enabling faster convergence and efficient repair.
}

\input{image/stat_num_steps.tex}

%% file: table/exe_time.tex
\begin{table}[t]
   \centering
   \caption{RQ2 - Execution time for each method on C100 and TinyImg.
   Row-wise ranking of $T_{\text{tot}}$ is highlighted with Gold (1st), Silver (2nd), and Bronze (3rd). 
   Abbreviations: Rep = \reptran, Ara = \arachne, Rand$_R$ = Random$_R$, Rand$_A$ = Random$_A$.
   }
   \small
   \setlength{\tabcolsep}{4pt}
   \renewcommand{\arraystretch}{1.0}
   \resizebox{\columnwidth}{!}{
   \begin{tabular}{ll*{2}{c}*{2}{c}|*{4}{c}}
     \toprule
     \multirow{2}{*}{Dataset} & \multirow{2}{*}{Rank / Type} &
     \multicolumn{2}{c}{$T_{\text{select}}$ [sec.]} &
     \multicolumn{2}{c}{$T_{\text{repair}}$ [sec.]} &
     \multicolumn{4}{c}{$T_{\text{tot}}$ [sec.]}\\
     \cmidrule(lr){3-4}\cmidrule(lr){5-6}\cmidrule(lr){7-10}
        &  &
       Rep & Ara &
       Rep & Ara &
       Rep & Ara & Rand$_R$ & Rand$_A$\\
     \midrule
 \multirow{9}{*}{C100}
  & 1 / \textit{SRC-TGT} & 21.06 & \textbf{18.55} & \textbf{329.99} & 405.18 &
     \cellcolor{silver}351.04 & 423.72 & \cellcolor{bronze}418.875 & \cellcolor{gold}241.674\\
  & 1 / \textit{TGT-FP} & \textbf{11.73} & 13.86 & \textbf{386.89} & 407.92 &
     \cellcolor{silver}398.61 & \cellcolor{bronze}421.78 & 423.179 & \cellcolor{gold}221.351\\
  & 1 / \textit{TGT-FN} & \textbf{11.66} & 17.90 & \textbf{281.87} & 408.97 &
     \cellcolor{gold}293.53 & 426.87 & \cellcolor{bronze}414.403 & \cellcolor{silver}341.957\\ \cmidrule{2-10}
  & 2 / \textit{SRC-TGT} & \textbf{11.68} & 16.94 & \textbf{264.59} & 404.16 &
     \cellcolor{gold}276.27 & 421.09 & \cellcolor{bronze}417.089 & \cellcolor{silver}293.511\\
  & 2 / \textit{TGT-FP} & \textbf{11.78} & 15.55 & \textbf{331.96} & 417.26 &
     \cellcolor{silver}343.74 & 432.81 & \cellcolor{bronze}386.160 & \cellcolor{gold}273.072\\
  & 2 / \textit{TGT-FN} & \textbf{11.68} & 17.06 & \textbf{287.50} & 418.63 &
     \cellcolor{gold}299.18 & 435.69 & \cellcolor{bronze}413.794 & \cellcolor{silver}307.170\\ \cmidrule{2-10}
  & 3 / \textit{SRC-TGT} & \textbf{11.76} & 15.83 & \textbf{296.74} & 404.27 &
     \cellcolor{silver}308.50 & 420.10 & \cellcolor{bronze}409.712 & \cellcolor{gold}280.190\\
  & 3 / \textit{TGT-FP} & \textbf{11.73} & 17.32 & \textbf{332.61} & 414.86 &
     \cellcolor{silver}344.34 & 432.18 & \cellcolor{bronze}386.064 & \cellcolor{gold}237.540\\
  & 3 / \textit{TGT-FN} & \textbf{11.74} & 17.17 & \textbf{370.24} & 403.76 &
     \cellcolor{silver}381.98 & \cellcolor{bronze}420.93 & 421.981 & \cellcolor{gold}283.828\\
 \midrule
 \multirow{9}{*}{TinyImg}
  & 1 / \textit{SRC-TGT} & \textbf{32.73} & 35.00 & \textbf{511.41} & 808.01 &
     \cellcolor{gold}544.14 & 843.01 & \cellcolor{bronze}820.283 & \cellcolor{silver}633.284\\
  & 1 / \textit{TGT-FP} & \textbf{21.60} & 24.25 & \textbf{646.00} & 841.71 &
     \cellcolor{silver}667.60 & 865.96 & \cellcolor{bronze}858.033 & \cellcolor{gold}473.249\\
  & 1 / \textit{TGT-FN} & \textbf{21.14} & 23.02 & \textbf{539.31} & 816.44 &
     \cellcolor{gold}560.45 & 839.47 & \cellcolor{bronze}780.018 & \cellcolor{silver}621.619\\ \cmidrule{2-10}
  & 2 / \textit{SRC-TGT} & \textbf{21.01} & 27.55 & \textbf{626.97} & 708.07 &
     \cellcolor{silver}647.98 & \cellcolor{bronze}735.61 & 824.027 & \cellcolor{gold}528.644\\
  & 2 / \textit{TGT-FP} & \textbf{21.30} & 22.79 & \textbf{597.41} & 826.76 &
     \cellcolor{gold}618.70 & 849.56 & \cellcolor{bronze}773.139 & \cellcolor{silver}655.785\\
  & 2 / \textit{TGT-FN} & \textbf{21.18} & 24.27 & \textbf{728.61} & 744.41 &
     \cellcolor{silver}749.79 & \cellcolor{bronze}768.67 & 773.695 & \cellcolor{gold}618.982\\ \cmidrule{2-10}
  & 3 / \textit{SRC-TGT} & \textbf{21.08} & 23.71 & \textbf{347.78} & 808.91 &
     \cellcolor{gold}368.85 & 832.62 & \cellcolor{bronze}821.061 & \cellcolor{silver}507.497\\
  & 3 / \textit{TGT-FP} & \textbf{21.17} & 23.31 & \textbf{640.50} & 790.13 &
     \cellcolor{gold}661.67 & 813.44 & \cellcolor{bronze}806.136 & \cellcolor{silver}711.400\\
  & 3 / \textit{TGT-FN} & \textbf{21.09} & 23.47 & \textbf{733.97} & 769.84 &
     \cellcolor{silver}755.06 & \cellcolor{bronze}793.31 & 820.926 & \cellcolor{gold}519.099\\ \midrule
  Average & & 17.62 & 20.97 & 458.57 & 599.96 &
     476.19 & 620.93 & 609.37 & 430.547\\
     \bottomrule
 \end{tabular}
 }
 \label{tab:exec_time}
 \end{table}

%% file: table/stat_exe_time.tex
\begin{table}[t]
\centering
\small
\caption{
RQ2 - Wilcoxon signed-rank tests on $T_{\text{tot}}$.
Each cell reports Cliff's $\delta$ together with statistical significance.
Significance levels: $^{\ast}p{<}.05$, $^{\ast\ast}p{<}.01$.
Abbreviations: Rep = \textsc{RepTran}, Ara = \arachne, Rand$_R$ = Random$_R$, Rand$_A$ = Random$_A$.
}
\label{tab:stat_exec_time}
\resizebox{0.9\columnwidth}{!}{
\begin{tabular}{lllllll}
\toprule
Dataset &
\makecell{Rep\\vs.\ Ara} &
\makecell{Rep\\vs.\ Rand$_R$} &
\makecell{Rep\\vs.\ Rand$_A$} &
\makecell{Ara\\vs.\ Rand$_R$} &
\makecell{Ara\\vs.\ Rand$_A$} &
\makecell{Rand$_R$\\vs.\ Rand$_A$} \\
\midrule
C100     & $-$1.00 $^{\ast}$ & $-$1.00 $^{\ast}$ & $+$0.33 & $-$0.33 & $+$1.00 $^{\ast}$ & $+$1.00 $^{\ast}$ \\
TinyImg  & $-$1.00 $^{\ast}$ & $-$1.00 $^{\ast}$ & $-$0.11 & $-$0.56 & $+$1.00 $^{\ast}$ & $+$1.00 $^{\ast}$ \\
\bottomrule
\end{tabular}
}
\end{table}

%% file: image/stat_num_steps.tex
\begin{figure}[t]
    \begin{center}
      \includegraphics[width=.8\linewidth]{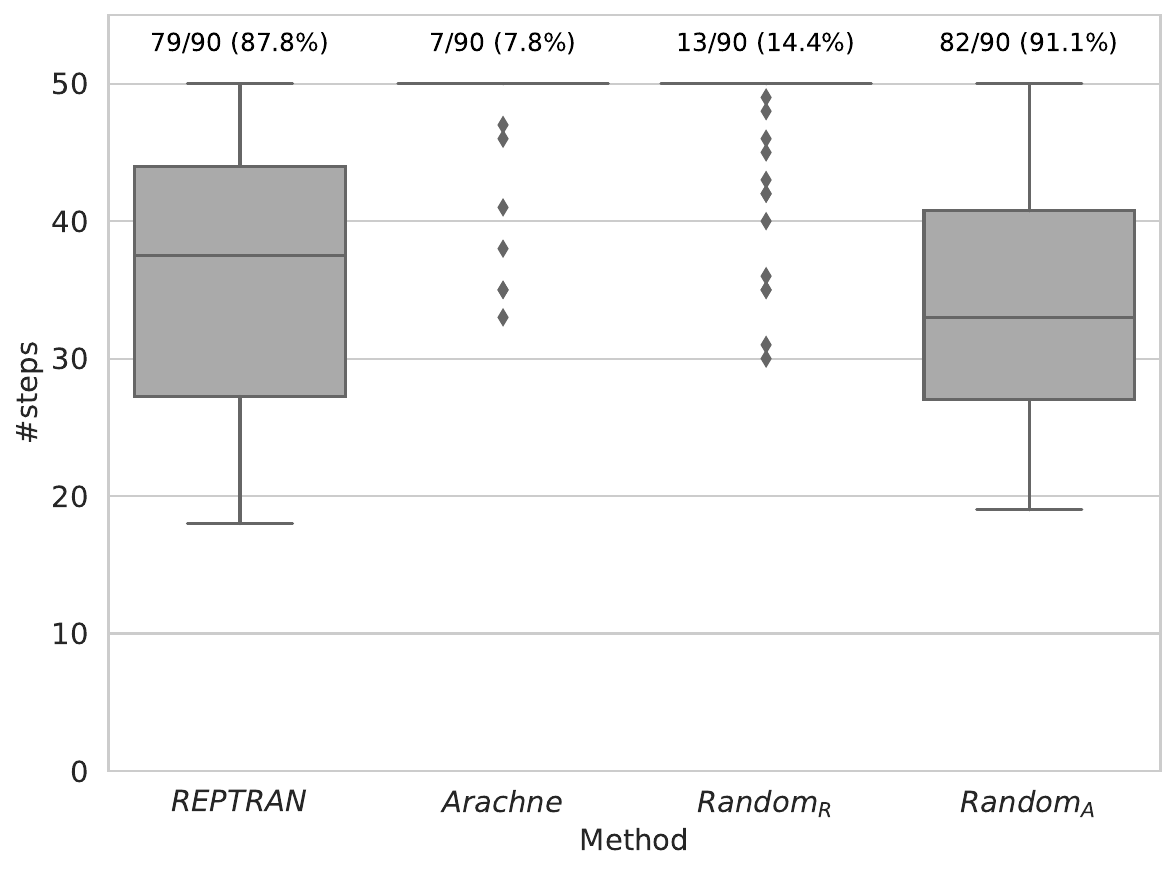}
      \caption{
        RQ2 - Number of steps taken in the search phase for each method ($G_{\text{max}}=50$, $G_{\text{stag}}=10$).
        Each box represents the distribution over 90 data points (2 datasets $\times$ 3 misclassification types $\times$ 3 ranks $\times$ 5 runs).
        The numbers above the boxes indicate the frequency of early stopping.
        }
      % 各ボックスは手法ごとのサーチフェーズのステップ数を示しており，データセット2 x 誤分類タイプ3 x ランク3 x 実行5 = 90個のデータ点からなります．
      \label{fig:num_steps}
    \end{center}
  \end{figure}

%% file: section/rqs/rq3.tex
\subsection{RQ3: Effect of the Number of Selected Weights}

\input{image/rr_br_scatter_per_num_weight.tex}

\textbf{Approach.}
The number of selected weights can significantly impact the outcome.
Selecting too few weights may lead to insufficient repair, whereas selecting too many may introduce unintended side effects.
To investigate this trade-off, we conducted experiments by varying the number of selected weights: 11 (the average number selected by \arachne), 236, 472, and 944 (corresponding to 0.005\%, 0.01\%, and 0.02\% of all weights, respectively), as described in \secref{sec:exp_config}.
To enable control over the number of selected weights in \arachne, we implement a variant called \arachnew, which selects the top-$N_w$ weights with the highest average score of forward impact and gradient loss, instead of relying on the Pareto front.

\textbf{Results.}
\figref{fig:rr_br_scatter} shows the repair and break rates for each number of selected weights.
To directly evaluate the trade-off between repair and break rates, we compute the harmonic mean of $RR$ and $1 - BR$.\footnote{Since lower break rates are better, we use the complement ($1 - \text{break rate}$) so that higher values indicate better performance, similar to the repair rate.}
The resulting scores are summarized in \tabref{tab:rr_br_trade_off}.

\textbf{Improvements in repair rate are often accompanied by an increase in break rate.}
From \figref{fig:rr_br_scatter}, both \textsc{RepTran} and \arachnew show higher repair rate when more than 236 weights are modified, but this typically results in a rise in break rate as well.
While some configurations (especially on TinyImg) achieve favorable combinations of high repair rate and low break rate, the overall tendency reveals a trade-off between maximizing repair effectiveness and minimizing unintended side effects.

\textbf{The trade-off between the repair and break rates is most effectively balanced at $N_w = 944$, but the improvement saturates after $N_w = 236$, indicating a diminishing benefit from selecting more weights.}
As shown in the rightmost column of \tabref{tab:rr_br_trade_off}, the average harmonic mean reaches its maximum at $N_w = 944$, suggesting that this setting achieves the best balance between improving the repair rate and minimizing the break rate.
However, increasing $N_w$ from 236 to 472 or 944 results in only marginal gains in the harmonic mean.
% \figref{fig:rr_br_scatter} also shows that the repair and break rates show little change between $N_w = 236$, $472$, and $944$.
This suggests that selecting more than 236 weights provides limited additional benefit in optimizing the trade-off.

\tabref{tab:rr_br_values} shows the results of our statistical comparison of repair and break rates, as in the previous RQs.
Statistically significant differences ($p < .05$) accompanied by a large positive effect size ($\delta > 0.474$) are highlighted in yellow.
For all weight settings, \textsc{RepTran} and \arachnew consistently produced statistically significantly higher repair and break rates than the random baseline, with all comparisons showing large effect sizes and $p < .05$.

\textbf{\textsc{RepTran} statistically significantly outperforms \arachnew in only one out of eight comparisons for the repair rate (C100 with 11 weights).}
However, seven out of eight cases show a positive effect size, with large effects observed when the number of selected weights is small (\eg, for C100, $\delta = +0.67$ with 11 weights and $\delta = +0.56$ with 236 weights).
This indicates a general tendency for \textsc{RepTran} to yield higher repair rates than \arachnew, though the statistical support is limited.

\summaryblock{Answer to RQ3}{
While 944 weights achieved the best trade-off between repair and break rates, improvements largely saturated at 236, suggesting limited benefit from adding more.
\textsc{RepTran} generally outperformed \arachnew in the repair rate, showing positive effect sizes in seven out of eight comparisons.
}

\input{table/rr_br_trade_off.tex}

\input{table/stat_test_perf_per_weight.tex}

%% file: image/rr_br_scatter_per_num_weight.tex
\begin{figure}[t]
  \centering
  \subfloat[C100]{%
    \includegraphics[width=.85\linewidth]{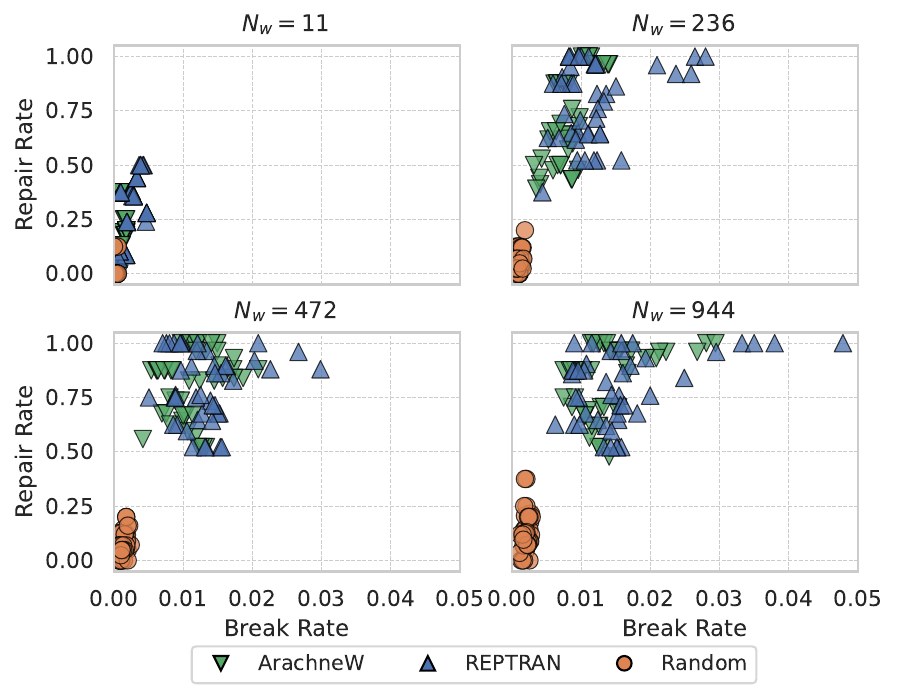}%
    \label{fig:rr_br_scatter_c100}%
  }\\ % 改行＋ちょっと余白
  \subfloat[TinyImg]{%
    \includegraphics[width=.85\linewidth]{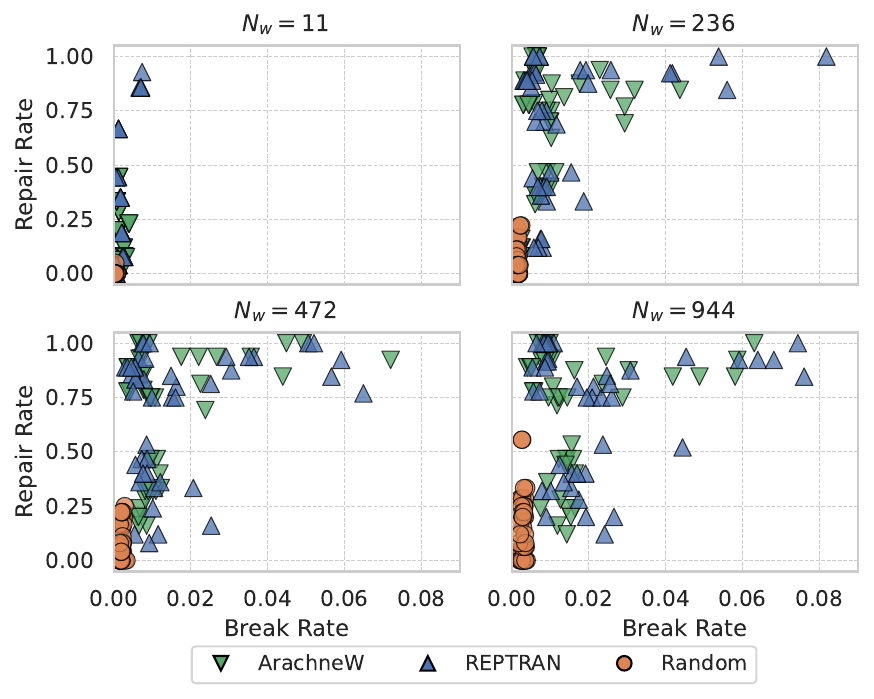}%
    \label{fig:rr_br_scatter_tiny}%
  }
  \caption{RQ3 - $RR$ and $BR$ for different methods and number of weights.
  % Each panel contains 135 points (3 methods $\times$ 3 types $\times$ 3 ranks $\times$ 5 runs).
  }
  \label{fig:rr_br_scatter}
\end{figure}

%% file: table/rr_br_trade_off.tex
\begin{table}[t]
  \centering
  \caption{RQ3 - Harmonic mean of $RR$ and 1 $-$ $BR$.
  % Bold values indicate the highest score among the three methods for each number of weights.
  }
  \scalebox{0.8}{
  \begin{tabular}{llcccc}
    \toprule
    Dataset & $N_w$ & \textsc{RepTran} & \arachnew & Random & Average \\
    \midrule
    \multirow{4}{*}{C100}
      & 11  & 0.494 & 0.263 & 0.017 & 0.258 \\
      % & 11  & \textbf{0.494} & 0.263 & 0.017 & 0.258 \\
      & 236 & 0.910 & 0.853 & 0.097 & 0.620 \\
      % & 236 & \textbf{0.910} & 0.853 & 0.097 & 0.620 \\
      & 472 & 0.914 & 0.908 & 0.141 & 0.655 \\
      % & 472 & \textbf{0.914} & 0.908 & 0.141 & 0.655 \\
      & 944 & 0.909 & 0.921 & 0.254 & 0.695 \\
      % & 944 & 0.909 & \textbf{0.921} & 0.254 & 0.695 \\
    \midrule
    \multirow{4}{*}{TinyImg}
      & 11  & 0.430 & 0.313 & 0.005 & 0.249 \\
      % & 11  & \textbf{0.430} & 0.313 & 0.005 & 0.249 \\
      & 236 & 0.810 & 0.789 & 0.118 & 0.572 \\
      % & 236 & \textbf{0.810} & 0.789 & 0.118 & 0.572 \\
      & 472 & 0.814 & 0.811 & 0.189 & 0.605 \\
      % & 472 & \textbf{0.814} & 0.811 & 0.189 & 0.605 \\
      & 944 & 0.819 & 0.822 & 0.276 & 0.639 \\
      % & 944 & 0.819 & \textbf{0.822} & 0.276 & 0.639 \\
    % \midrule
    % Average & 
    %   & 0.763 & 0.710 & 0.137 & 0.537 \\
      % & \textbf{0.763} & 0.710 & 0.137 & 0.537 \\
    \bottomrule
  \end{tabular}
  }
  \label{tab:rr_br_trade_off}
\end{table}

%% file: table/stat_test_perf_per_weight.tex
\definecolor{strong}{RGB}{255,255,153}  % 薄い黄色（必要に応じて変更）
\definecolor{weak}{RGB}{200,215,255}
\begin{table}[t]
\centering
\caption{
RQ3 - Wilcoxon signed-rank tests on $RR$ and $BR$.
Each cell reports Cliff's $\delta$ together with statistical significance.
Significance levels: $^{\ast}p{<}.05$, $^{\ast\ast}p{<}.01$.
Abbreviations: Rep = \textsc{RepTran}, AraW = \arachnew, Rand = Random.
}
\label{tab:rr_br_values}
\resizebox{0.9\columnwidth}{!}{
\begin{tabular}{ll*{3}{c}*{3}{c}}
\toprule
\multirow{3}{*}{Dataset} & \multirow{3}{*}{$N_w$} &
\multicolumn{3}{c}{$RR$} & \multicolumn{3}{c}{$BR$} \\
\cmidrule(lr){3-5}\cmidrule(lr){6-8}
 &  & \footnotesize\makecell{Rep vs.\\ AraW} & \footnotesize\makecell{Rep vs.\\ Rand} & \footnotesize\makecell{AraW vs.\\ Rand} & 
         \footnotesize\makecell{Rep vs.\\ AraW} & \footnotesize\makecell{Rep vs.\\ Rand} & \footnotesize\makecell{AraW vs.\\ Rand} \\
\midrule
% -------------------- C100 --------------------
\multirow{4}{*}{C100}
   & 11  & \cellcolor{strong}+0.67$^*$ & \cellcolor{strong}+1.00$^*$ & \cellcolor{strong}+0.89$^*$ & \cellcolor{strong}+1.00$^*$ & \cellcolor{strong}+1.00$^*$ & \cellcolor{strong}+1.00$^*$ \\
   & 236 & +0.56 & \cellcolor{strong}+1.00$^*$ & \cellcolor{strong}+1.00$^*$ & +0.33 & \cellcolor{strong}+1.00$^*$ & \cellcolor{strong}+1.00$^*$ \\
   & 472 & +0.11 & \cellcolor{strong}+1.00$^*$ & \cellcolor{strong}+1.00$^*$ & \cellcolor{strong}+0.56 & \cellcolor{strong}+1.00$^*$ & \cellcolor{strong}+1.00$^*$ \\
   & 944 & $-0.44$ & \cellcolor{strong}+1.00$^*$ & \cellcolor{strong}+1.00$^*$ & +0.33 & \cellcolor{strong}+1.00$^*$ & \cellcolor{strong}+1.00$^*$ \\ 

% -------------------- TinyImg --------------------
\midrule
\multirow{4}{*}{TinyImg}
   & 11  & +0.33 & \cellcolor{strong}+0.89$^*$ & \cellcolor{strong}+0.89$^*$ & $-0.11$ & \cellcolor{strong}+1.00$^*$ & \cellcolor{strong}+1.00$^*$ \\
   & 236 & +0.22 & \cellcolor{strong}+1.00$^*$ & \cellcolor{strong}+1.00$^*$ & \cellcolor{strong}+0.56 & \cellcolor{strong}+1.00$^*$ & \cellcolor{strong}+1.00$^*$ \\
   & 472 & +0.22 & \cellcolor{strong}+1.00$^*$ & \cellcolor{strong}+1.00$^*$ & +0.33 & \cellcolor{strong}+1.00$^*$ & \cellcolor{strong}+1.00$^*$ \\
   & 944 & +0.33 & \cellcolor{strong}+1.00$^*$ & \cellcolor{strong}+1.00$^*$ & \cellcolor{strong}+0.78$^*$ & \cellcolor{strong}+1.00$^*$ & \cellcolor{strong}+1.00$^*$ \\

\bottomrule
\end{tabular}}
\end{table}

%% file: section/rqs/rq4.tex
\subsection{RQ4: Effect of the Balance Coefficients}
\textbf{Approach.}
\textsc{RepTran} uses one balance coefficient in each phase.
The coefficient $p$ in the selection phase (Eq.~\ref{eq:weightsuspscore}) balances ModFI and ModGL when scoring weights, whereas $\alpha$ in the search phase (Eq.~\ref{eq:fitness}) balances maintaining correct behavior and reducing misbehavior.
% Since $\alpha$ controls how strongly the fitness function emphasizes misclassifications, higher values of $\alpha$ are expected to increase both the repair and break rates.
In this RQ, we analyze the effect of each coefficient in isolation by varying $p$ over \{0.1, 0.5, 0.9\} and $\alpha$ over \{1, 2, 4, 8, 10\}, following the same setup as in the original \arachne experiments for $\alpha$~\cite{sohn2022tosem}.
We fix the number of weights at $N_w = 236$, which showed a good balance in RQ3.
This allows us to eliminate the influence of the number of selected weights and measure the effect of each balance coefficient.

\textbf{Results.}
For space, we do not report the raw repair and break rates for each value of $p$ and $\alpha$ here; these are available in our replication package, and we focus on the results of the statistical tests.
We conducted a Kruskal--Wallis test~\cite{kruskal1952test} to determine whether the distributions of repair or break rates differ across the tested values of each balance coefficient.
The null hypothesis for each dataset, metric, and coefficient states that all groups for the coefficient come from the same distribution, \ie, the repair or break rates are identically distributed regardless of the coefficient value.

\textbf{The selection-phase coefficient $p$ did not significantly affect repair or break rates.}
Across the tested values of $p$, no clear trend emerged in either repair or break rates.
As shown in \tabref{tab:kw_results}, all null hypotheses were retained for both datasets and both metrics.
This result indicates that the choice of $p$ has little impact on the repair outcome.

\textbf{No statistically significant differences were observed in the repair or break rates across different values of $\alpha$.}
As shown in \tabref{tab:kw_results}, all null hypotheses were retained for $\alpha$, with $p$-values approaching 1.0 in every case.
This indicates not only a lack of statistically significant differences across $\alpha$ values, but also that the observed repair and break rates are highly consistent with the assumption of identical distributions.
These findings further support the conclusion that the balance coefficient $\alpha$ does not have a meaningful impact on repair or break rates.

% \figref{fig:alpha_plot} shows the repair and break rates (y-axis) for different values of $\alpha$ (x-axis).
% Contrary to expectations, these rates do not exhibit a clear monotonic increase as $\alpha$ increases.
% Even when a slight upward trend is observed, its impact is limited.
% Taken together with the findings from RQ3, this suggests that the number of selected weights ($N_w$) in the selection phase has a greater impact on the repair result than the balance coefficients $p$ and $\alpha$.

\summaryblock{Answer to RQ4}{
Varying the balance coefficients $p$ from 0.1 to 0.9 and $\alpha$ from 1 to 10 had minimal impact on repair and break rates.
Kruskal--Wallis tests showed no statistically significant differences for either coefficient, indicating that $p$ and $\alpha$ do not meaningfully affect the outcome.}

\input{table/stat_alpha.tex}

%% file: table/stat_alpha.tex
\begin{table}[t]
  \centering
  \small
  \caption{RQ4 - Kruskal--Wallis test results for $RR$ and $BR$ across different balance-coefficient values.}
  \label{tab:kw_results}
  \scalebox{0.8}{
  \begin{tabular}{lcccc}
  \toprule
    & \multicolumn{2}{c}{$p$ for selection phase} & \multicolumn{2}{c}{$\alpha$ for search phase} \\
  \cmidrule(lr){2-3} \cmidrule(lr){4-5}
  Dataset & $RR$ $p$-value & $BR$ $p$-value & $RR$ $p$-value & $BR$ $p$-value \\
  \midrule
  C100    & 0.404 & 0.614 & 0.996 & 0.984 \\
  TinyImg & 0.997 & 0.468 & 0.994 & 0.939 \\
  \bottomrule
  \end{tabular}
  }
  \end{table}

%% file: section/rqs/rq5.tex
\subsection{RQ5: Comparison with a ViT-specific Method}
\label{sec:rq5}

\textbf{Approach.}
We compare \textsc{RepTran} with the three PRoViT variants, all targeting the same final-block FFN.
% The key contrast is the size of the edit: \textsc{RepTran} modifies only 472 weights (0.01\% of the FFN parameters), whereas the PRoViT variants optimize the entire weights parameters in the FFN ($\sim$4.72M weight parameters).
We use the same 18 benchmarks as in previous RQs and report each metric as the percentage change $(\text{PRoViT}/\textsc{RepTran}-1)\times100$, where a positive change means PRoViT repairs more (better), breaks more (worse), or is slower (worse), respectively.
We set a time budget of 1{,}800 seconds for solving the LP and treat any run exceeding it as a timeout (TO).
This budget is generous, as it is more than twice the largest repair time observed for any method in RQ2, which stays below 1{,}000 seconds (\tabref{tab:exec_time}).

\input{table/provit_results.tex}

\textbf{Results.}
\tabref{tab:provit_results} reports how each PRoViT variant compares with \textsc{RepTran} on repair rate ($RR$), break rate ($BR$), and repair time ($T_{\text{tot}}$).
A \textcolor{ForestGreen}{green} (\textcolor{red}{red}) cell indicates that the variant is better (worse) than \textsc{RepTran} on that metric, and the bottom row aggregates these changes over the benchmarks.

\textbf{On average, no PRoViT variant improves over \textsc{RepTran} on all three metrics at once.}
As the Average row shows, each variant is worse than \textsc{RepTran} on at least one metric: PRoViT$_\mathrm{LP}$ on both break rate ($+22\%$) and time ($+90\%$), PRoViT$_\mathrm{FT}$ on break rate ($+20\%$), and PRoViT$_\mathrm{FT+LP}$ on time ($+51\%$).
That is, the gains in repair rate that the variants achieve come at the cost of either more broken behaviors or substantially longer repair time.

\textbf{PRoViT$_\mathrm{LP}$ and PRoViT$_\mathrm{FT+LP}$ are slower than \textsc{RepTran} and do not scale to larger label spaces.}
These two variants are on average $+90\%$ and $+51\%$ slower, as both are bottlenecked by solving the LP.
This cost grows with the label space.
On TinyImg the slowdown reaches up to $+203\%$, and PRoViT$_\mathrm{LP}$ fails to return a solution in 5 of the 9 TinyImg benchmarks.
\textsc{RepTran} has no such scalability wall, since it performs a localized repair rather than solving a global LP.

\textbf{PRoViT$_\mathrm{FT}$ is the only variant faster than \textsc{RepTran}, but at the cost of overfitting.}
It fine-tunes until the few dozen samples in the repair set (at most 56, as shown in \tabref{tab:fault_bench_size}) are fully fixed, which explains both its speed ($-84\%$) and its higher break rate ($+20\%$).
Fixing the repair set does not by itself guarantee generalization to unseen inputs.

\summaryblock{Answer to RQ5}{
No PRoViT variant improves over \textsc{RepTran} on all of repair rate, break rate, and time on average: each is worse on at least one.
\textsc{RepTran} attains comparable repair and break rates while running $51$ to $90\%$ faster than the LP-based variants, one of which (PRoViT$_\mathrm{LP}$) times out on 5 of the 9 TinyImg benchmarks; the only faster variant, PRoViT$_\mathrm{FT}$, achieves its speed by overfitting the repair set.
}

%% file: table/provit_results.tex
\begin{table}[t]
  \centering
  \caption{RQ5 - Repair rate ($RR$), break rate ($BR$), and repair time ($T_{\text{tot}}$) of the three PRoViT variants relative to \textsc{RepTran}.
  Arrows $\uparrow$ ($\downarrow$) indicate that higher (lower) values are better.
  TO denotes a time out.}
  \label{tab:provit_results}
  \definecolor{greycell}{HTML}{C0C0C0}
  \providecommand{\gd}[1]{\textcolor{ForestGreen}{#1}}
  \providecommand{\bd}[1]{\textcolor{red}{#1}}
  \setlength{\tabcolsep}{3pt}
  \resizebox{\columnwidth}{!}{
  \begin{tabular}{l*{3}{ccc}}
  \toprule
  \multirow{2}{*}{Dataset / Rank / Type} &
  \multicolumn{3}{c}{PRoViT$_\mathrm{LP}$} &
  \multicolumn{3}{c}{PRoViT$_\mathrm{FT}$} &
  \multicolumn{3}{c}{PRoViT$_\mathrm{FT+LP}$}\\
  \cmidrule(lr){2-4}\cmidrule(lr){5-7}\cmidrule(lr){8-10}
   & $RR\uparrow$ & $BR\downarrow$ & $T_{\text{tot}}\downarrow$ & $RR\uparrow$ & $BR\downarrow$ & $T_{\text{tot}}\downarrow$ & $RR\uparrow$ & $BR\downarrow$ & $T_{\text{tot}}\downarrow$\\
  \midrule
   C100 / 1 / \textit{SRC-TGT} & \gd{$+5\%$}  & \bd{$+28\%$} & \gd{$-48\%$}  & \gd{$+5\%$}  & \bd{$+28\%$}  & \gd{$-100\%$} & \gd{$+5\%$}  & \bd{$+16\%$} & \gd{$-46\%$}\\
   C100 / 1 / \textit{TGT-FP}  & \bd{$-18\%$} & \bd{$+9\%$}  & \bd{$+279\%$} & \bd{$-18\%$} & \gd{$-18\%$}  & \gd{$-99\%$}  & \bd{$-22\%$} & \gd{$-23\%$} & \bd{$+55\%$}\\
   C100 / 1 / \textit{TGT-FN}  & \gd{$+6\%$}  & \bd{$+77\%$} & \bd{$+23\%$}  & \gd{$+6\%$}  & \bd{$+145\%$} & \gd{$-100\%$} & \gd{$+6\%$}  & \bd{$+87\%$} & \bd{$+8\%$}\\ \cmidrule{1-10}
   C100 / 2 / \textit{SRC-TGT} & \gd{$+9\%$}  & \gd{$-20\%$} & \gd{$-8\%$}   & \gd{$+25\%$} & \bd{$+18\%$}  & \gd{$-100\%$} & \gd{$+9\%$}  & \gd{$-37\%$} & \bd{$+16\%$}\\
   C100 / 2 / \textit{TGT-FP}  & \gd{$+8\%$}  & \gd{$-35\%$} & \bd{$+163\%$} & \gd{$+8\%$}  & \gd{$-22\%$}  & \gd{$-99\%$}  & \bd{$-2\%$}  & \gd{$-37\%$} & \bd{$+136\%$}\\
   C100 / 2 / \textit{TGT-FN}  & \gd{$+8\%$}  & \gd{$-17\%$} & \bd{$+221\%$} & \gd{$+8\%$}  & \bd{$+43\%$}  & \gd{$-100\%$} & \gd{$+8\%$}  & \gd{$-14\%$} & \bd{$+75\%$}\\ \cmidrule{1-10}
   C100 / 3 / \textit{SRC-TGT} & \gd{$+21\%$} & \gd{$-16\%$} & \gd{$-47\%$}  & \gd{$+21\%$} & \bd{$+8\%$}   & \gd{$-100\%$} & \gd{$+21\%$} & \gd{$-10\%$} & \gd{$-24\%$}\\
   C100 / 3 / \textit{TGT-FP}  & \bd{$-13\%$} & \bd{$+54\%$} & \bd{$+98\%$}  & \gd{$+8\%$}  & \gd{$-5\%$}   & \gd{$-97\%$}  & \bd{$-17\%$} & \bd{$+9\%$}  & \bd{$+113\%$}\\
   C100 / 3 / \textit{TGT-FN}  & \bd{$-1\%$}  & \gd{$-29\%$} & \bd{$+102\%$} & \gd{$+11\%$} & \bd{$+34\%$}  & \gd{$-100\%$} & \gd{$+1\%$}  & \gd{$-25\%$} & \bd{$+79\%$}\\
  \midrule
   TinyImg / 1 / \textit{SRC-TGT} & \multicolumn{3}{c}{\cellcolor{greycell}TO} & \gd{$+9\%$}  & \gd{$-4\%$}  & \gd{$-100\%$} & \gd{$+9\%$}  & \gd{$-4\%$}  & \gd{$-100\%$}\\
   TinyImg / 1 / \textit{TGT-FP}  & \multicolumn{3}{c}{\cellcolor{greycell}TO} & \gd{$+67\%$} & \gd{$-11\%$} & \gd{$-99\%$}  & \gd{$+50\%$} & \gd{$-59\%$} & \bd{$+182\%$}\\
   TinyImg / 1 / \textit{TGT-FN}  & \multicolumn{3}{c}{\cellcolor{greycell}TO} & \bd{$-1\%$}  & \gd{$-41\%$} & \gd{$-100\%$} & \bd{$-4\%$}  & \gd{$-47\%$} & \gd{$-68\%$}\\ \cmidrule{1-10}
   TinyImg / 2 / \textit{SRC-TGT} & \gd{$+11\%$} & \bd{$+26\%$} & \bd{$+18\%$} & \gd{$+11\%$} & \gd{$-8\%$}  & \gd{$-100\%$} & \gd{$+11\%$} & \gd{$-15\%$} & \bd{$+28\%$}\\
   TinyImg / 2 / \textit{TGT-FP}  & \multicolumn{3}{c}{\cellcolor{greycell}TO} & \gd{$+9\%$}  & \gd{$-8\%$}  & \gd{$-100\%$} & \gd{$+19\%$} & \gd{$-67\%$} & \bd{$+203\%$}\\
   TinyImg / 2 / \textit{TGT-FN}  & \multicolumn{3}{c}{\cellcolor{greycell}TO} & \gd{$+10\%$} & \bd{$+27\%$} & \gd{$-100\%$} & \bd{$-3\%$}  & \gd{$-54\%$} & \bd{$+48\%$}\\ \cmidrule{1-10}
   TinyImg / 3 / \textit{SRC-TGT} & \gd{$+3\%$}  & \gd{$-39\%$} & \bd{$+57\%$}  & \gd{$+3\%$}  & \gd{$-20\%$} & \gd{$-100\%$} & \gd{$+3\%$}  & \gd{$-20\%$} & \gd{$-100\%$}\\
   TinyImg / 3 / \textit{TGT-FP}  & \bd{$-39\%$} & \bd{$+85\%$} & \bd{$+173\%$} & $+0\%$       & \bd{$+218\%$}& \bd{$+172\%$} & \bd{$-4\%$}  & \gd{$-1\%$}  & \bd{$+160\%$}\\
   TinyImg / 3 / \textit{TGT-FN}  & \gd{$+11\%$} & \bd{$+166\%$}& \bd{$+137\%$} & \gd{$+4\%$}  & \gd{$-15\%$} & \gd{$-100\%$} & \bd{$-3\%$}  & \gd{$-67\%$} & \bd{$+146\%$}\\ \midrule
   \textbf{Average} & \gd{\boldmath$+1\%$} & \bd{\boldmath$+22\%$} & \bd{\boldmath$+90\%$} & \gd{\boldmath$+10\%$} & \bd{\boldmath$+20\%$} & \gd{\boldmath$-84\%$} & \gd{\boldmath$+5\%$} & \gd{\boldmath$-20\%$} & \bd{\boldmath$+51\%$}\\
   \bottomrule
  \end{tabular}
  }
\end{table}

%% file: section/discussion.tex
\section{Discussion} \label{sec:discussion}

\input{table/neuron_score_ablation.tex}

\subsection{Contribution of the Neuron Score Components}
As described in \secref{sec:neuron_score}, the neuron score is the product of two components, \textit{VDiff} and \textit{MisAct}.
To justify our proposed neuron score, we isolate the contribution of each component by ablating it at a fixed budget of $N_w = 236$, which showed a good balance between repairs and breaks in RQ3.
Keeping all else equal, we compare \textit{Full} (VDiff $\times$ MisAct), \textit{VDiff-only}, and \textit{MisAct-only} over the 18 fault benchmarks.
For each dataset, we statistically compare the $RR$ and $BR$ of Full against those of each ablated variant (VDiff-only and MisAct-only) using paired Wilcoxon signed-rank tests over the 45 paired values (9 benchmarks $\times$ 5 runs, Holm-corrected).
\tabref{tab:neuron_ablation} reports the resulting Cliff's $\delta$, where a positive $\delta$ means Full attains a higher $RR$ (or $BR$) than the ablated variant.

\textbf{MisAct is the dominant component, while VDiff is complementary.}
Removing MisAct (VDiff-only) significantly lowers $RR$ on TinyImg ($\delta = 0.42$, $p < .001$) and also on C100 ($\delta = 0.18$, not significant).
Moreover, as the signs of $\delta$ in \tabref{tab:neuron_ablation} show, no ablated variant attains both a higher $RR$ and a lower $BR$ than Full; any gain on one metric is offset by a loss on the other.
Combining the two components therefore yields a reasonably balanced selection across both datasets.

\subsection{How \textsc{RepTran} Differs from \arachnew and PRoViT?}
Although \textsc{RepTran} and \arachnew showed no statistically significant differences in most RQ3 settings, and RQ5 showed that no PRoViT variant dominates \textsc{RepTran} across all metrics, understanding how these methods differ is important for assessing their complementarity.
% We therefore examined (1) the differences in the weights selected by \textsc{RepTran} and \arachnew, and (2) the specific samples that \textsc{RepTran}, \arachnew, and PRoViT$_\mathrm{FT+LP}$ repair and break.

% Understanding how \textsc{RepTran} and \arachnew differ in their weight selection strategies is essential for assessing the unique contributions of our neuron-aware repair approach.
% While both \textsc{RepTran} and \arachnew aim to identify important weights for repairing Transformer models, they rely on different criteria.
% \arachnew selects weights solely based on bidirectional scores that evaluate the contribution of each weight.
% In contrast, \textsc{RepTran} incorporates an additional \emph{neuron score} to capture internal behavioral differences of the FFNs in Transformer models.

% \input{image/discussion_suspw.tex}
% \input{image/discussion_venn.tex}

\subsubsection{Differences in the selected weights}
We computed the Jaccard coefficients between the weights selected by \textsc{RepTran} and \arachnew across various configurations.
The Jaccard coefficient quantifies the degree of overlap between two sets, ranging from 0 (no overlap) to 1 (complete overlap).

\textbf{The median Jaccard coefficient across all configurations was 0.13, ranging from 0.00 to 0.47, indicating that the two methods often selected different weights.}
The highest coefficients were 0.47 for TinyImg, Rank 1, and \textit{TGT-FP}, suggesting that the methods may converge on similar weights for some misbehaviors.
Overall, however, the low overlap supports that incorporating neuron scores in \textsc{RepTran} yields weight selections qualitatively different from \arachnew.

\subsubsection{Differences in repaired and broken samples}
To examine whether the methods repair and break different samples, we compared the sets of test samples that \textsc{RepTran}, \arachnew, and PRoViT$_\mathrm{FT+LP}$ consistently repaired or broke across all five runs.
Among the three PRoViT variants, we focus on PRoViT$_\mathrm{FT+LP}$ because it is the best-performing one in RQ5.
We focus on such consistently repaired or broken samples to capture the inputs that each method reliably repairs or breaks rather than incidental cases.
For each method, we count the \emph{unique} samples that only that method repairs (or breaks) while the other two do not.
\tabref{tab:repaired_broken_cases} reports these counts, summed over the three ranks for each dataset and misclassification type.

\input{table/unique_rb.tex}

\textbf{While some misclassifications are repaired only by PRoViT$_\mathrm{FT+LP}$, far more inputs are broken only by it.}
It has the largest number of unique repairs in every row, but the counts are small.
At the same time, it also produces by far the most unique breaks (48--199 per row), several times more than \textsc{RepTran} or \arachnew and far exceeding its own unique repairs.
Thus, the few inputs that it alone repairs come at the cost of many more inputs that it alone breaks.

\textbf{\textsc{RepTran} and \arachnew behave far more conservatively, repairing and breaking few samples uniquely.}
This is because they modify only a localized subset of the FFN weights, whereas PRoViT$_\mathrm{FT+LP}$ optimizes the entire FFN and therefore alters the model's behavior more aggressively.
The only exception is TinyImg / \textit{TGT-FN}, where \textsc{RepTran} records the most unique breaks (101 samples), driven by a single benchmark (rank 2) rather than the general trend.

\subsection{Future Directions}
% \textsc{RepTran} shows that modifying a small number of weights in the FFNs of Transformer models can effectively repair specific misbehaviors.
While \textsc{RepTran} is promising as a practical repair method for Transformer models, it also leaves several important questions open for future research.
% our findings suggest several directions for future research.

\begin{comment}
% This design is motivated by both theoretical insights (e.g., prior work~\cite{geva2021emnlp,dai2022ACL,tang2024ACL,niu2024ICLR} highlighting the semantic role of FFNs) and empirical evidence (our analysis in \figref{fig:neuron_scores_grid}).
% However, it remains several limitations and opens promising directions for future work.
\end{comment}

\subsubsection{Applications to Large Models}
In light of recent advances in large models such as LLMs, investigating how to adapt \textsc{RepTran} for such models is a promising direction for future research.
These models are based on the Transformer architecture with billions of parameters and support widely used applications such as ChatGPT.
As model size increases, data collection and retraining become increasingly costly; thus, repairing specific misbehaviors with limited data and time is especially valuable for large models.
Since FFN components are essential even in state-of-the-art large models, \textsc{RepTran} is in principle applicable to them.
However, it remains an open question whether FFN-only repair remains effective at this scale and whether the selection and search phases can be executed within a practical time.

\subsubsection{Other Applications of Our Proposed Neuron Score}
While our neuron score is primarily designed to guide model repair, it also holds potential for broader quality assurance tasks beyond repair, such as testing~\citednntesting.
Because the score reflects neuron behavior in Transformer models, it may facilitate the development of testing strategies specifically tailored to Transformer-based architectures, including LLMs.
For example, test inputs can be synthesized to strongly activate neurons with high scores, helping to expose specific failures that may otherwise go undetected.
In addition, existing test inputs can be prioritized based on the activation patterns of high-scoring neurons, enabling more efficient detection of misbehaviors under limited testing resources.

%% file: table/neuron_score_ablation.tex
\begin{table}[t]
\centering
\caption{
Component ablation of the neuron score.
Significance levels: $^{\ast}p{<}.05$, $^{\ast\ast}p{<}.01$, $^{\ast\ast\ast}p{<}.001$.
}
\label{tab:neuron_ablation}
\resizebox{.8\columnwidth}{!}{
\begin{tabular}{llcc}
\toprule
Dataset & Contrast & $RR$ & $BR$ \\
\midrule
\multirow{2}{*}{C100}
   & Full vs.\ VDiff-only  & $+0.18$ & $-0.22$ \\
   & Full vs.\ MisAct-only & $-0.18$ & $-0.11$ \\
\midrule
\multirow{2}{*}{TinyImg}
   & Full vs.\ VDiff-only  & $+0.42^{\ast\ast\ast}$ & $+0.49^{\ast\ast}$ \\
   & Full vs.\ MisAct-only & $+0.07$ & $+0.13$ \\
\bottomrule
\end{tabular}}
\end{table}

%% file: table/unique_rb.tex
\begin{table}[t]
    \centering
    \caption{Number of unique repairs and breaks.
    Each cell lists counts in the order \textsc{RepTran}/\arachnew/PRoViT$_\mathrm{FT+LP}$; bold values indicate the highest count in that cell.}
    \label{tab:repaired_broken_cases}
    \setlength{\tabcolsep}{3pt}
    \scalebox{0.9}{
    \begin{tabular}{lcc}
    \toprule
    Dataset / Type & Unique Repairs & Unique Breaks \\
    \midrule
    C100 / \textit{SRC-TGT} & 0/0/\textbf{1} & 4/3/\textbf{49} \\
    C100 / \textit{TGT-FN} & 0/0/\textbf{8} & 19/7/\textbf{199} \\
    C100 / \textit{TGT-FP} & 4/4/\textbf{12} & 4/11/\textbf{105} \\ \midrule
    TinyImg / \textit{SRC-TGT} & 0/0/\textbf{2} & 3/2/\textbf{48} \\
    TinyImg / \textit{TGT-FN} & 1/0/\textbf{3} & \textbf{101}/22/57 \\
    TinyImg / \textit{TGT-FP} & 0/1/\textbf{5} & 6/36/\textbf{67} \\
    \bottomrule
    \end{tabular}}
\end{table}

%% file: section/validity.tex
\section{Threats to Validity} \label{sec:validity}
% We discuss potential threats to validity of our study along three dimensions: internal, external, and construct threats.

\textbf{Internal Threats.}
\textsc{RepTran} involves several hyperparameters, such as the number of selected weights $N_w$, the balance coefficients $p$ and $\alpha$, and those for differential evolution.
While these values were chosen based on prior studies or preliminary experiments, alternative settings may yield different outcomes.
Although RQ3 and RQ4 investigate the effects of some key hyperparameters (i.e., $N_w$, $p$, and $\alpha$), a comprehensive hyperparameter search is not included.
Therefore, our findings may be partially dependent on specific configurations.

\textbf{External Threats.}
We evaluated \textsc{RepTran} on two image classification datasets using ViT.
To ensure coverage of diverse misclassifications, we constructed 18 fault benchmarks by varying misclassification types and ranks.
Moreover, in all experiments, the samples used for repair and those used for evaluation were strictly separated to simulate realistic deployment scenarios.
While these design choices enhance the generalizability of our findings, it remains unclear whether the results extend to other domains or architectures.

\textbf{Construct Threats.}
We define successful repair as the correction of targeted misclassifications without breaking existing correct predictions, which aligns with prior work on DNN repair~\cite{tokui2022saner,sohn2022tosem,nawas2024isaiv}.
However, this definition may not fully capture broader quality attributes such as robustness, fairness, or safety, which are also important in practical deployments.
Prior work~\cite{you2025arxiv,ishimoto2025tosem} suggests that the optimal repair method may vary depending on the specific quality attribute being targeted.
Therefore, developing and evaluating repair methods for Transformer models from multiple perspectives beyond correctness remains an important direction for future work.

%% file: section/conclusion.tex
\section{Conclusion} \label{sec:conclusion}
This paper introduced \reptran, a search-based repair method tailored to Transformer models.
We evaluated \reptran on 18 fault benchmarks derived from CIFAR-100 and Tiny-ImageNet.
The results showed that \reptran statistically significantly outperforms both random selection and \arachne in repair rate.
It also slightly outperforms \arachnew, as indicated by the effect size.
While \reptran incurs a higher break rate than some baselines, this rate remains below 5.7\% and is justified by its substantially higher repair effectiveness.
Our analysis in the discussion further showed that \reptran has complementary strengths to \arachnew and PRoViT.
These findings highlight the effectiveness of \reptran and demonstrate that search-based repair can be successfully extended beyond CNNs and RNNs to modern Transformer architectures.

%% file: section/data_availability.tex
\section{Data Availability}
\label{sec:data_availability}

All data and code used in this study are available at the following repository:
\url{https://github.com/posl/RepTran-replication}.